\documentclass[lettersize,journal]{IEEEtran}
\usepackage{amsmath,amsfonts,amssymb}
\usepackage{algorithmic}
\usepackage{array}
\usepackage[caption=false,font=normalsize,labelfont=sf,textfont=sf]{subfig}
\usepackage{textcomp}
\usepackage{stfloats}
\usepackage{url}
\usepackage{verbatim}
\usepackage{graphicx}
\usepackage{cite}
\usepackage{mathtools}
\usepackage{bigdelim}
\usepackage{blkarray}
\usepackage{multicol}
\usepackage{hyperref}

\usepackage[thinc]{esdiff}

\usepackage{diagbox}
\usepackage{orcidlink}
\usepackage{bbm}
\usepackage{algorithm}

\DeclareMathOperator{\argmin}{argmin}
\DeclareMathOperator{\argmax}{argmax}
\DeclareMathOperator{\VAR}{\textsc{Var}}     
\DeclareMathOperator{\predictionstep}{\textsc{PredictionStep}}     
\DeclareMathOperator{\updatestep}{\textsc{UpdateStep}}     
\DeclareMathOperator{\chol}{\textsc{Chol}}     
\DeclareMathOperator{\MSE}{MSE}    
\DeclareMathOperator{\sequence}{\textsc{Seq}}

\DeclareMathOperator{\cqpoints}{\textsc{CQpoints}}
\DeclareMathOperator{\reward}{\textsc{Reward}}

\DeclareMathOperator{\voi}{\textsc{VoI}}

\usepackage{amsthm}

\newcommand{\algorithmictraining}{\textbf{Training:}}
\newcommand\TRAINING{\item[\algorithmictraining]}
\newcommand{\algorithmictesting}{\textbf{Testing:}}
\newcommand\TESTING{\item[\algorithmictesting]}

\newcommand{\ceil}[1]{\lceil {#1} \rceil}
\newcommand{\floor}[1]{\lfloor {#1} \rfloor}

\usepackage{mathrsfs}
\usepackage{fge}

\usepackage{nameref}

\newcounter{mylabelcounter}

\makeatletter
\newcommand{\labelText}[2]{%
#1\refstepcounter{mylabelcounter}%
\immediate\write\@auxout{%
  \string\newlabel{#2}{{1}{\thepage}{{\unexpanded{#1}}}{mylabelcounter.\number\value{mylabelcounter}}{}}%
}%
}
\makeatother

\begin{document}

\title{Goal-Oriented Sensor Reporting Scheduling for Non-linear Dynamic System Monitoring}

\author{Prasoon Raghuwanshi\orcidlink{0000-0002-9629-9742},~\IEEEmembership{Student Member,~IEEE},
Onel Luis Alcaraz López\orcidlink{0000-0003-1838-5183},~\IEEEmembership{Senior Member,~IEEE},
I-Hong Hou\orcidlink{0000-0002-1166-8773},~\IEEEmembership{Senior Member,~IEEE},
Vimal Bhatia\orcidlink{0000-0001-5148-6643},~\IEEEmembership{Senior Member,~IEEE},
Matti Latva-aho\orcidlink{0000-0002-6261-0969},~\IEEEmembership{Fellow,~IEEE}
\thanks{Prasoon Raghuwanshi, Onel Luis Alcaraz López, and Matti Latva-aho are with the Centre for Wireless Communications, University of Oulu, $90570$, Oulu, Finland (e-mail: Prasoon.Raghuwanshi@oulu.fi; Onel.AlcarazLopez@oulu.fi; Matti.Latva-aho@oulu.fi).}
\thanks{I-Hong Hou is with the Department of Electrical and Computer Engineering,	Texas A$\&$M University, $77840$, College Station, Texas, United States (e-mail: ihou@tamu.edu)}
\thanks{Vimal Bhatia is with the Department of Electrical Engineering, Indian Institute of Technology Indore, $453552$, Indore, India, with the Skoda Auto University, $29301$, Mlada Boleslav, Czech Republic, with the Faculty of Informatics and Management, University of Hradec Krolove, $50003$, Hradec Krolove, Czechia, and with the University of Oulu, $90570$, Oulu, Finland (e-mail: vbhatia@iiti.ac.in)}
\thanks{This research has been supported by the Research Council of Finland (Grants 362782 (ECO-LITE), and 369116 (6G Flagship)), the Finnish Foundation for Technology Promotion, the INDIFICORE project (Grant 24650101111), the European Commission through the Horizon Europe/JU SNS project Hexa-X-II (Grant 101095759), the Riitta ja Jorma J. Takasen säätiö (Grant 20240358), the Nokia Scholarship (Grant 20260695), and the Oulun yliopiston tukisäätiö (Grant 20260126).}}

\maketitle

\begin{abstract}
Goal-oriented communication (GoC) is a form of semantic communication where the effectiveness of information transmission is measured by its impact on achieving the desired goal.
In Internet-of-Things (IoT) networks, GoC can enable sensors to selectively transmit data relevant to \textcolor{black}{the} intended goals of the receiver, thereby facilitating timely decision-making, reducing network congestion, and enhancing spectral efficiency.
In this paper, we consider an IoT scenario where an edge node polls sensors monitoring the state of a non-linear dynamic system (NLDS) to respond to the queries of several clients.
This work delves into the foregoing GoC problem and solution, which we termed goal-oriented scheduling (GoS).
The latter utilizes deep reinforcement learning (DRL) with meticulously devised action space, state space, and reward function.
A long short-term memory network is used to estimate the inter-query duration and the corresponding estimation standard deviation.
This empowers the proposed DRL scheduler to make judicious decisions, even when no queries are posed.
Numerical analysis demonstrates that the proposed GoS \textcolor{black}{reduces the mean square error (MSE) of the query responses compared to the benchmark scheduling methods even as the number of clients and DRL action space increase, which proves its scalability.}
Moreover, this is attained without polling sensors during \textcolor{black}{$\mathbf{70\% - 87\%}$} of the testing phase, \textcolor{black}{thus promoting} energy efficiency.
\textcolor{black}{Lastly, the complexity of the proposed GoS is relatively lower than the benchmark scheduling methods.}
\end{abstract}


\section{Introduction}\label{intro_section1}

\IEEEPARstart{T}{here} are billions of Internet-of-Things (IoT) devices worldwide, and their number will keep on growing exponentially in the coming years \cite{10285066}.
Notably, a significant share of the IoT landscape comprises low-cost/low-power sensors monitoring dynamic systems, which usually have high-dimensional states.
As a result, \textcolor{black}{a massive amount of data is} increasingly exchanged in IoT communications, often under stringent quality of service, e.g., latency and reliability, requirements \cite{10364354, 10012472, 10443958}.

Given the resource limitations inherent to the IoT sensors and networks, there has \textcolor{black}{been growing} interest in remotely estimating system states at a fusion center/edge node\textcolor{black}{\cite{8812616, 9437348, 9593268, 10409276, 10156782, 9551200, 10323428, 11523609, 10_1145_3565287_3610263, cao2023goal, luo2023goal, elessawy2026real, 11389914, 11475799, 9036854, 9768131, 10143239, 10273599}.}
Notably, an edge node may remotely estimate the entire system state by gathering observations from a subset of IoT sensors, rather than the entire sensor network\textcolor{black}{, thus} ultimately resulting in energy-efficient state observation.
The application of remote state estimation (RSE) assisted-sensor reporting scheduling is diverse, spanning fields such as voltage regulation in power systems \cite{7479120}, strategic actuator placement in control systems \cite{tzoumas2015minimal}, and sensing/reporting scheduling in wireless networks \cite{9036854, 9768131, 10143239}.

\subsection{State-of-art on RSE-assisted Sensor Reporting Scheduling: Markovian \textcolor{black}{and Linear} System}\label{intro_subsection1_1}

Recently,\textcolor{black}{\cite{8812616, 9437348, 9593268, 10409276, 10156782, 9551200, 10323428, 11523609, 10_1145_3565287_3610263, cao2023goal, luo2023goal, elessawy2026real, 11389914, 11475799}} explored RSE-assisted sensor reporting scheduling.
\textcolor{black}{These studies perform RSE of the Markovian process \cite{8812616, 9437348, 9593268, 10409276, 10156782, 9551200, 10323428, 10_1145_3565287_3610263, cao2023goal, luo2023goal, elessawy2026real, 11389914, 11475799} and/or linear dynamic process \cite{cao2023goal, 11523609} at an edge node.}
Examples of dynamic systems that can be described through the Markovian processes include control systems such as node mobility in ad-hoc networks, unmanned aerial vehicle systems, and robotic swarms, \textcolor{black}{as well as} physical processes such as transfer of a liquid/gas \cite{9437348}.
Meanwhile, frequency deviation in the load frequency control system can be described through a linear dynamic process \cite{cao2023goal}.
In\textcolor{black}{\cite{8812616, 9437348, 9593268, 10409276, 10156782, 9551200, 10323428, 11523609, 11389914, 11475799}}, an agent, which may be an IoT device, monitors a single process, while \textcolor{black}{an} agent monitors multiple processes in\textcolor{black}{\cite{10_1145_3565287_3610263, cao2023goal, luo2023goal, elessawy2026real}.}
The agent schedules sampling times of the observed process(es) and forwards the sampled observation to the edge node over a wireless channel to perform RSE.
\textcolor{black}{Note that \cite{11523609} considers a single agent and a channel that wears out due to aging and transmission tasks; \cite{8812616, 9437348, 9593268, 10409276, 10156782, 9551200, 10323428, 10_1145_3565287_3610263, cao2023goal, luo2023goal, 11389914, 11475799} consider non-wearing channels and follow the notion of single packet reception; and \cite{elessawy2026real} considers a non-wearing and multi-packet reception channel.}
Later, the edge node sends an acknowledgement message to the agent, consisting of information about the recently completed RSE task.
The objective in\textcolor{black}{\cite{8812616, 9437348, 9593268, 10409276, 10_1145_3565287_3610263, 10156782, 9551200, cao2023goal, luo2023goal, elessawy2026real, 11389914, 11475799, 10323428, 11523609}} is to derive a scheduling strategy for the aforesaid sampling times that
(i) \textcolor{black}{minimizes} the RSE error at the edge node\textcolor{black}{\cite{8812616, 9437348, 9593268, 10409276, 10_1145_3565287_3610263, 11523609}};
(ii) \textcolor{black}{minimizes} the cost of actuation error, which encapsulates the significance of RSE error at the time of actuation\textcolor{black}{\cite{10156782, 9551200, luo2023goal, 10323428, elessawy2026real}};
(iii) \textcolor{black}{minimizes} the probability that the state of the linear dynamic process violates its preset values \cite{cao2023goal};
\textcolor{black}{(iv) achieves a balanced tradeoff between the cost of prolonged error in RSE at the edge node and the transmission cost \cite{11389914}; and
(v) minimizes the cost of prolonged error in RSE at the edge node\cite{11475799}.}

{
\setlength\arrayrulewidth{1pt}
\begin{table}[!t]
\caption{Examples of Client Queries \label{querytypes_table}}
\centering
\begin{tabular}{@{}l l@{}}
\hline
\textbf{Query} & \textbf{Definition, $z_{c}(\mathbf{x}(t))$}  \\ 
\hline
Current state & $\mathbf{x}(t)$  \\
Maximum component & $\max(\mathbf{x}(t))$  \\
Count range & $\sum_{m=1}^{M} \mathbbm{1}(x_{m}(t) \in [a,  b])$  \\
Sample mean & $\frac{1}{M} \sum_{m=1}^{M} x_{m}(t)$ \\
Sample variance & $\frac{1}{M-1} \sum_{m=1}^{M} (x_{m}(t) - \frac{1}{M} \sum_{m=1}^{M} x_{m}(t))^2$  \\
\hline
\multicolumn{2}{l}{$^{\ast}$ Herein, ${\mathbf{x}(t) = [ x_1(t), \cdots, x_M(t) ]^T \in \mathbb{R}^{M}}$.}\\
\end{tabular}
\end{table}
}

Notably,\textcolor{black}{\cite{8812616, 9437348, 9593268, 10409276, 10_1145_3565287_3610263, 10156782, 9551200, cao2023goal, luo2023goal, elessawy2026real, 11389914, 11475799, 10323428, 11523609} derived scheduling strategies on the agent side, leading to an increase in its energy consumption,} which is not desired in scenarios where agents are low-power IoT devices.
Moreover, these studies are only valid for the RSE of either the Markovian or linear dynamic process.
In practical applications, however, a process to be remotely estimated can be \textcolor{black}{a} non-Markovian or non-linear dynamic.
Additionally, \cite{9437348, 8812616} assume an error-free wireless \textcolor{black}{channel;}\textcolor{black}{\cite{10156782, cao2023goal, elessawy2026real, 11389914, 11475799, 11523609}} assume that the wireless channel statistics are known to the agent; \textcolor{black}{and \cite{8812616, 9437348, 9593268, 10409276, 10_1145_3565287_3610263, 10156782, 9551200, cao2023goal, luo2023goal, elessawy2026real, 11389914, 11475799, 10323428, 11523609} assume that the edge node's remote state estimates are known to the agent.}
Meanwhile, the low-complexity scheduling policy proposed in \cite{luo2023goal} is only valid for the case when the wireless channel statistics and the observed process statistics are known to the agent.
While \cite{luo2023goal} also proposes a deep reinforcement learning (DRL)-based scheduling policy for the case when channel/process statistics are unknown, however, the high computational complexity of this policy makes it unsuitable for implementation on IoT devices.
Furthermore, \cite{9036854, 9768131, 10143239} state that, at the time of sampling, sensors typically taint their sampled observations.
This tainting process can be expressed through an analytical expression consisting of an observation model and measurement noise \cite{9036854, 9768131, 10143239}.
Here, the observation model determines how much of the system's state is directly observable from measurement, while the measurement noise accounts for real-world uncertainties, sensor inaccuracies, or external disturbances affecting the measurement.
\textcolor{black}{Nevertheless}, \textcolor{black}{\cite{8812616, 9437348, 10409276, 10_1145_3565287_3610263, 10156782, 9551200, cao2023goal, luo2023goal, elessawy2026real, 11389914, 11475799, 10323428}} assume that no such tainting is involved at the time of sampling, \textcolor{black}{while} the inclusion of measurement noise in \cite{9593268} changed the structure of samples from Markovian to non-Markovian, leading \cite{9593268} to derive a non-optimal scheduling strategy.
\textcolor{black}{Lastly, \cite{10_1145_3565287_3610263, 11523609}} investigates the signal-agnostic RSE, where the RSE error is expressed as a function of the \textcolor{black}{age-of-information} (AoI) \cite{9358178, 10323422} metric.
In such settings,\textcolor{black}{\cite{10_1145_3565287_3610263, 11523609}} concludes that the scheduling problem for minimizing the RSE error transforms into an AoI minimization problem.
However, the AoI metric can only partially characterize the variations in system state, \textcolor{black}{while} the AoI-based scheduling policy that aims to minimize the AoI of the observed system does not minimize its RSE error \cite{8812616, 9437348}.
\textcolor{black}{Moreover,} AoI-based performance metrics/scheduling policies \cite{9358178} overlook the potential mismatch between the content required at the edge node with respect to a particular task/goal and the content delivered to the edge node\textcolor{black}{\cite{10156782, 9551200, 10195938, elessawy2026real, 11389914}}.
These limitations have prompted the research community to go beyond the notion of AoI in the case of RSE assisted-sensor reporting scheduling.

\begin{figure}[!t]
\centering
\includegraphics[width=\linewidth]{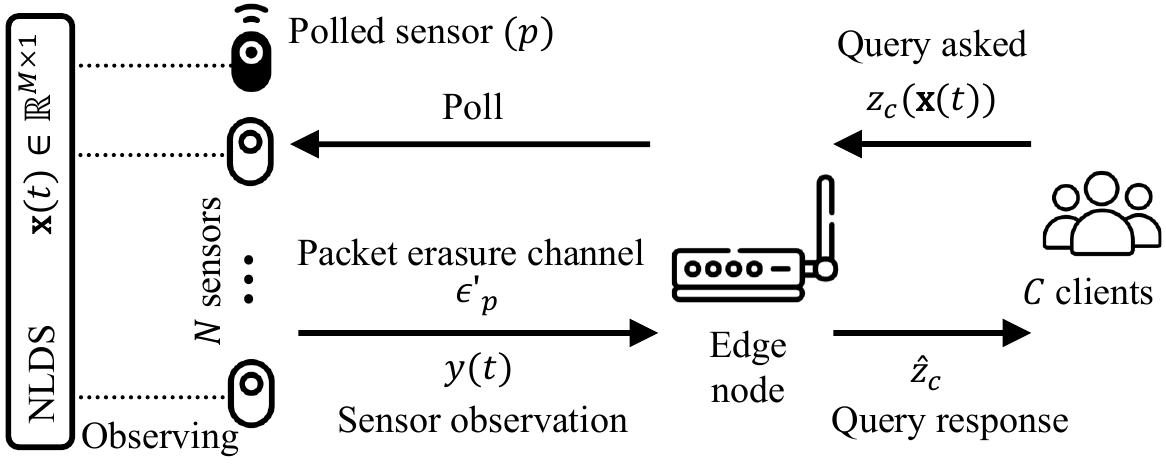}
\caption{GoC illustration.
Clients ask queries to the edge node about the non-linear dynamic system state observed by sensors with the non-linear observation model.
The edge node, based on the decision taken by its scheduler, may poll a sensor.
Besides, the edge node responds to queries based on the state estimate computed by the cubature quadrature Kalman filter.}
\label{systemmodelfigure}
\end{figure}

\subsection{State-of-art on VoI-based RSE-assisted Sensor Reporting Scheduling}\label{sectionI_B}

Value-of-information ($\voi$) has been suggested in \cite{10143239, 10273599, li2023towards} as a suitable metric for quantifying the impact of sensor transmission on the RSE error.
Here, RSE error is defined with respect to the desired \textit{goal}.
A goal might be to accurately identify the system state, or even to accurately respond to queries \cite{10884945, 10976624} from clients in systems like the one illustrated in Fig.~\ref{systemmodelfigure}.
In such \textcolor{black}{a} goal-oriented communication (GoC) system, clients pose queries to the edge node in order to obtain information about the system state, observed by sensors.
Table~\ref{querytypes_table} provides examples of potential client queries.

Recently, authors in \cite{9036854, 9768131, 10143239} utilized VoI-based RSE-assisted sensor reporting scheduling at an edge node.
The objective in \cite{9036854} is to identify the state of a linear dynamic system, whereas \textcolor{black}{the focus in \cite{9768131, 10143239}} is on effectively addressing client queries regarding the state of a linear dynamic system.
Thus, the $\voi$ adopted in \cite{9036854} corresponds to the mean square error $(\MSE)$ of the state estimation.
Meanwhile, $\voi$ is defined in \cite{9768131, 10143239} as the difference between the $\MSE$ of query response relative to the prior and posterior estimates of the state estimator \cite{10273599}.
Here, prior and posterior estimates denote estimates obtained before and after the sensor transmission, respectively.
Furthermore, \cite{9036854, 9768131, 10143239} exploit a key advantage offered by RSE, \textcolor{black}{namely the} ability to observe system states by selectively polling a subset of sensors.
In \cite{9036854}, the sensor scheduling strategy is devised to minimize the state estimation $\MSE$, whereas in\textcolor{black}{\cite{9768131, 10143239} it} aims to minimize the $\MSE$ of the query response.

Interestingly, a closed-form mathematical expression for the $\MSE$ of the query response, with respect to the posterior estimates, can be obtained for certain queries like sample mean, sample variance, and current state.
Thus, sensor reporting scheduling strategies for such queries can be determined analytically, as depicted in \cite{9768131}. 
Conversely, for queries such as the maximum component and count range of the system state, deriving closed-form mathematical expressions for the $\MSE$ of the query response, with respect to the posterior estimates, proves to be unattainable.
Therefore, addressing such queries necessitates the utilization of advanced approaches such as DRL to tackle the sensor reporting scheduling problem, as outlined in \cite{10143239}.

Note that proposals in \cite{9036854, 9768131, 10143239} have one common limitation: they only work for a linear dynamic system model and linear observation model, a prerequisite for Kalman filter-based RSE as the Kalman filter cannot deal with a non-linear dynamic system (NLDS) and non-linear observation model (NLOM).
Besides, in \cite{10143239}, a sensor must be polled at every time step, even when there are no client queries, resulting in unnecessary depletion of sensor energy.
Apart from that, the complete state of the Kalman filter is provided as input in \cite{10143239} to its DRL-based sensor scheduler.
This input significantly inflates the size of the deep neural network (DNN) utilized by the DRL-based sensor scheduler, as it must also extract relevant information from the input.
On top of that, time instances where no queries are posed are treated uniformly, providing the same reward to the DRL-based sensor scheduling algorithm on all those time instances.
Consequently, the proposal in \cite{10143239} struggles in the absence of queries.
Finally, the sensor scheduler in \cite{10143239} also demands information about the state of the query process, at the client, to anticipate the time duration after which the next query would be asked.
Indeed, the query process is modeled through memoryless Markov chains, wherein the time elapsed since the last query from the client is sufficient to determine the current state of the query process \cite{10143239}.
This does not hold for a \textit{memory-dependent} query process (MQP), and hence, the proposals in \cite{10143239} cannot be applied here.

\subsection{Contribution}

We address the aforesaid limitations regarding NLDS, NLOM, MQP, and RSE-assisted-sensor reporting scheduling in the current work, while proposing a novel sensor reporting scheduling approach.
For this, we consider the GoC system in Fig.~\ref{systemmodelfigure}, wherein the edge node orchestrates sensor reporting scheduling to gather partial yet informative sensor observations to respond to potential queries from clients.
Herein, the query generation process at the clients has memory, contrary to \cite{10143239}.
Our specific contributions in this paper are as follows:
\begin{itemize}
    \item We propose a novel sensor reporting scheduling approach termed \textit{goal-oriented scheduling} (GoS) for IoT sensors with NLOM tasked with monitoring NLDS.
    The sole motive of sensor reporting scheduling is to minimize the $\MSE$ of future query responses, hence the term \textit{goal-oriented} scheduling.
    \item We devise a long short-term memory (LSTM) network to estimate the inter-query duration and the corresponding estimation standard deviation.
    \item We devise an action space and a reward function for a DRL-based sensor scheduler such that it does not poll a sensor at every time step.
    Meanwhile, the state space is designed to involve the attributes estimated by the LSTM and the NLDS state estimated using the cubature quadrature Kalman filter (CQKF) \cite{phdthesis}.
    The devised state space and reward function empower our scheduler to make judicious decisions, even when no queries are posed, which would later lead to the minimization of the $\MSE$ of the query response.
    \item We show that the proposed scheduler exhibits the least complexity compared to the scheduler from \cite{10143239} and the Monte Carlo scheduler \cite{9768131}, both used as benchmarks.
    Moreover, the numerical results reveal that our proposed scheduler obtains the least $\MSE$ of the query response.
    Besides, this is accomplished by reducing the number of sensor transmissions, avoiding sensor polling during \textcolor{black}{$70\% - 87\%$} of the testing phase, thereby saving valuable energy resource.
\end{itemize}

\subsection{Organization}\label{subsec_organizations}

The paper is structured as follows.
Section~\ref{systemmodelsection} delineates the system model.
Section~\ref{goalorientedschedulingsection} describes the components of the GoS framework and presents the scheduling problem.
Section~\ref{benchmarkschedulersection} introduces benchmark schedulers and Section~\ref{sectionschedulerComplexities} discusses the computational complexities of all the considered schedulers.
Section~\ref{resultssection} presents the numerical results.
Lastly, Section~\ref{conclusionsection} concludes the paper and outlines potential avenues for future research.

\textit{Notation}: ${\argmax(\cdot)}$ and \textcolor{black}{${\argmin(\cdot)}$ denote the argument of the maximum and minimum function, while ${\max(\cdot)}$ denotes} the maximum function.
The cardinality of a set is represented by ${|\cdot|}$, while the transpose operation is denoted by ${[\cdot]^T}$.
Column vectors/matrices are indicated by boldface lowercase/uppercase letters.
The determinant, ceil function, floor function, and the expected value are denoted by ${\det(\cdot)}$, $\ceil{\cdot}$, $\floor{\cdot}$, and ${\mathbb{E}[\cdot]}$, respectively.
${\mathbf{I}_M}$ and ${\mathbf{0}_M}$ signify the ${M \times M}$ identity matrix and null vector of dimension $M$, respectively.
Additionally, ${\mathbf{1}_{p}}$ denotes a vector with all elements set to zero except the $p^{th}$ element, which is $1$.
The sets ${\mathbb{R}^{M}}$, ${\mathbb{R}_{+}^{M}}$, ${\mathbb{R}_{\geq0}^{M}}$, and ${\mathbb{N}^{M}}$ represent a real, real positive, real non-negative, and non-negative integer vector, respectively, all with dimension $M$, while set ${\mathbb{R}^{M \times M}}$ represent real matrix of dimension ${M \times M}$.
A Gaussian sample vector with mean ${\mathbf{\bar{y}}}$ and covariance matrix ${\mathbf{Z}}$ is denoted as ${\mathbf{y} \sim \mathcal{N}(\mathbf{\bar{y}},\mathbf{Z})}$.
The indicator function, Cholesky decomposition \textcolor{black}{\cite{6710599}}, sample variance, and uniform distribution between $0$ and $1$ are denoted by ${\mathbbm{1}(\cdot)}$, $\chol(\cdot)$, ${\VAR(\cdot)}$, and ${\mathcal{U}(0,1)}$, respectively.

\section{System Model} \label{systemmodelsection}

Consider the GoC system illustrated in Fig.~\ref{systemmodelfigure}, where an edge node receives data from $N$ sensors indexed by ${n \in \{ 1, 2, \cdots, N \}}$ and \textcolor{black}{responds} to queries from a set $\mathscr{C}$ of $C$ remote clients.
A query from client ${c \in \mathscr{C}}$ is a request for the value of the function ${z_{c}(\mathbf{x}(t))}$, while the edge node responds to it with an estimate ${\hat{z}_{c}}$.
Each client asks a different type of query about the system state.
The system operates in discrete time slots, labeled as ${t}$.
In each slot, the edge node decides whether to poll a single sensor or refrain from doing so.
The state of the NLDS is
\begin{align}
    \mathbf{x}(t) = \mathbf{f}(\mathbf{x}(t-1)) + \mathbf{v}(t) \in \mathbb{R}^{M},
\end{align}
where $M$ is the dimensionality of the NLDS state, ${\mathbf{f}(\cdot)}$ represents a \textit{non-linear state dynamics} (NLSD) \textit{function}, and $\mathbf{v}(t) \sim \mathcal{N}(\mathbf{0}_M, \mathbf{\Sigma})$ denotes the Gaussian noise with zero mean and covariance ${\mathbf{\Sigma} \in \mathbb{R}^{M \times M}}$.
Meanwhile, sensors observe the system state as captured by
\begin{align}
    \mathbf{y}(t) = \mathbf{g}(\mathbf{x}(t)) + \mathbf{v}'(t) \in \mathbb{R}^{N},
\end{align}
\textcolor{black}{where} $\mathbf{g}(\cdot)$ represents a \textit{non-linear observation function}, and $\mathbf{v}'(t) \sim \mathcal{N}(\mathbf{0}_N, \mathbf{\Sigma}')$ is the zero-mean Gaussian measurement noise with covariance matrix ${\mathbf{\Sigma}' \in \mathbb{R}^{N \times N}}$.
Additionally, we model the channel between the sensor $n$ and the edge node as a packet erasure channel with a transmission error probability $\epsilon'_{n}$.
\textcolor{black}{We} assume ${\mathbf{f}(\cdot)}$, ${\mathbf{g}(\cdot)}$, ${\mathbf{\Sigma}}$, and ${\mathbf{\Sigma}'}$ are known at the edge node.

\begin{algorithm}[!t]
\caption{$\cqpoints$}\label{CQpoints}
\algsetup{
linenosize=\small,
linenodelimiter=.}
\begin{algorithmic}[1]
\REQUIRE $M, n'$

\STATE Find the intersection points $\boldsymbol{\psi}_{j}, \forall j \in \{ 1, \cdots, 2M \}$ of the unit $M$-hyper-sphere and its axes    \hspace*{\fill} $\triangleright \ \boldsymbol{\psi}_{j} :$ cubature point

\STATE Compute the roots ${\upsilon_{j'}, \forall j' \in \{ 1, \cdots, n' \}}$ of the CL polynomial\hspace*{\fill} $\triangleright \ \upsilon_{j'} :$ quadrature point

\STATE \label{CQpointequation} $\boldsymbol{\xi}_{j' + (j-1)n'} = \sqrt{2 \upsilon_{j'}} \boldsymbol{\psi}_{j},$     \hspace*{\fill} $\triangleright$ CQ point \\
$w_{j' + (j-1)n'} = \frac{n'!}{2M} \frac{\Gamma(\ceil{\frac{M}{2} - 1} + n' + 1)}{\Gamma(M/2) \upsilon_{j'}} \frac{1}{L'(\upsilon_{j'})^2},$ \\
\hspace*{\fill}$\forall j \in \{1, \cdots, 2M \}, \forall j' \in \{1, \cdots, n' \}$

\ENSURE $\mathbf{w} = [w_{1}, \cdots, w_{2M n'}]^T, \boldsymbol{\Xi} = [\boldsymbol{\xi}_{1}, \cdots, \boldsymbol{\xi}_{2M n'}]^T$

\end{algorithmic}
\label{CQpoints}
\end{algorithm}

\begin{algorithm}[!t]
\caption{CQKF at $t$}\label{CQKFalgorithm}
\algsetup{
linenosize=\small,
linenodelimiter=.}
\begin{algorithmic}[1]
\REQUIRE $\hat{\mathbf{x}}(t-1), \hat{\mathbf{\Psi}}(t-1), \mathbf{\Sigma}, \mathbf{w}, \boldsymbol{\Xi}, \mathbf{\Sigma}', p$

\STATE \label{step1cqkfalgo} $\mathbf{x}'(t), \mathbf{\Psi}'(t), \{\boldsymbol{\zeta}_{i}(t-1)| \forall i \in \{1, \cdots, 2M n'\}\} \leftarrow$ \\
\hspace*{\fill} $\predictionstep(\hat{\mathbf{x}}(t-1), \hat{\mathbf{\Psi}}(t-1), \mathbf{\Sigma}, \mathbf{w}, \boldsymbol{\Xi})$

\STATE \label{step2cqkfalgo} Draw $\theta$ from $\mathcal{U}(0,1)$

\IF{$(p \mathrel{\ne} 0)$ \textbf{and} $(\theta  \geq 0.02 \ceil{\frac{p-1}{10}})$}
  \STATE $\hat{\mathbf{x}}(t), \hat{\mathbf{\Psi}}(t) \leftarrow \updatestep(\mathbf{x}'(t), \mathbf{\Psi}'(t), \mathbf{\Sigma}', \mathbf{w}, \boldsymbol{\Xi}, p,$ \\
  \hspace*{\fill} $y(t), \{\boldsymbol{\zeta}_{i}(t-1)| \forall i \in \{1, \cdots, 2M n'\}\})$
\ELSE
  \STATE $\{ \hat{\mathbf{x}}(t), \hat{\mathbf{\Psi}}(t) \} = \{ \mathbf{x}'(t), \mathbf{\Psi}'(t) \}$
\ENDIF \label{step7cqkfalgo}

\ENSURE $\mathbf{x}'(t), \mathbf{\Psi}'(t), \hat{\mathbf{x}}(t), \hat{\mathbf{\Psi}}(t)$

\end{algorithmic}
\label{CQKFalgorithm}
\end{algorithm}

\section{Goal-Oriented Scheduling} \label{goalorientedschedulingsection}

The proposed GoS framework comprises the following \textcolor{black}{four key components: state estimator, sensor scheduler, query process at the clients, and estimator for inter-query duration and its corresponding standard deviation. Detailed descriptions of each component and the training process for the sensor scheduler and inter-query duration estimator} are provided next.

\subsection{System State Estimator} \label{CQKFsubsection}

We employ CQKF for NLDS state estimation, as its estimates have relatively higher accuracy than the conventional filters for NLDS, such as the cubature Kalman filter and extended Kalman filter \cite{https://doi.org/10.1049/iet-spr.2012.0085}.
As initialization, CQKF requires cubature quadrature (CQ) points $(\boldsymbol{\Xi})$ and their corresponding weights $(\mathbf{w})$, whose computation procedure is described in Algorithm~\ref{CQpoints}. 
Initially, we determine the cubature points ${\boldsymbol{\psi}_{j}, \forall j \in \{ 1, \cdots, 2M \}}$, which are the intersection points of the unit $M$-hyper-sphere and its axes.
For example, the unit $2$-hyper-sphere, also known as the unit circle, has ${[1,0]^T, [0,1]^T, [-1,0]^T}$ and ${[0,-1]^T}$ as cubature points, which are basically the intersection points of the unit $2$-hyper-sphere with its axes.
Likewise, the unit $M$-hyper-sphere has ${\boldsymbol{\psi}_{j} = \mathbf{1}_{j}, \boldsymbol{\psi}_{M+j} = -\mathbf{1}_{j}, \forall j \in \{ 1, \cdots, M \}}$, as cubature points.
Subsequently, we compute the roots ${\upsilon_{j'}, \forall j' \in \{ 1, \cdots, n' \}}$ of the Chebyshev-Leguerre (CL) polynomial \cite{phdthesis, https://doi.org/10.1049/iet-spr.2012.0085}, known as quadrature points.
Here, the CL polynomial is given as
\begin{align}
        L(\upsilon) = & \sum_{k=0}^{n'} \binom{n'}{k} (-1)^{k} \frac{(n'+ \ceil{\frac{M}{2} - 1})!}{(n'+ \ceil{\frac{M}{2} - 1} -k)!} \upsilon^{n'-k} \nonumber \\
        = & \ell_{0} + \ell_{1} \upsilon + \cdots + \ell_{n'-1} \upsilon^{n'-1} + \upsilon^{n'}.
\end{align}
To compute quadrature points, we first have to formulate the companion matrix ${(\mathbf{D})}$ \cite{li2024companion}, corresponding to $L(\upsilon)$, as
\begin{align}
    \mathbf{D} = \begin{bmatrix}
                    \mathbf{0}_{n'-1} & \mathbf{I}_{n'-1} \\
                    -\ell_{0} & -\boldsymbol{\ell}^T
                \end{bmatrix}.
\end{align}
where $\boldsymbol{\ell} = [\ell_{1}, \cdots, \ell_{n'-1}]^T$.
Next, we formulate the characteristic polynomial \cite{li2024companion} of ${\mathbf{D}}$, which is ${\det(\mathbf{D} - \upsilon \mathbf{I}_{n'})}$, here $\upsilon$ corresponds to the eigenvalues of $\mathbf{D}$. Note that, ${L(\upsilon) = \det(\mathbf{D} - \upsilon \mathbf{I}_{n'})}$. Therefore, the eigenvalues of $\mathbf{D}$ are the roots of ${L(\upsilon)}$.
Finally, we determine $\boldsymbol{\Xi}$ and $\mathbf{w}$ by utilizing the cubature and quadrature points in step~\ref{CQpointequation} of Algorithm~\ref{CQpoints}, respectively.
Note that, ${L'(\upsilon_{j'})}$ in step~\ref{CQpointequation} of Algorithm~\ref{CQpoints} is the first derivative of ${L(\cdot)}$ at ${\upsilon = \upsilon_{j'}}$.

The CQKF is detailed in Algorithm~\ref{CQKFalgorithm} and encompasses two steps: \textit{prediction step} and \textit{update step}, elaborated thoroughly in Algorithm~\ref{predictionstepCQKF} and \ref{updatestepCQKF}, respectively.

\begin{algorithm}[!t]
\caption{$\predictionstep$}\label{predictionstepCQKF}
\algsetup{
linenosize=\small,
linenodelimiter=.}
\begin{algorithmic}[1]
\REQUIRE $\hat{\mathbf{x}}(t-1), \hat{\mathbf{\Psi}}(t-1), \mathbf{\Sigma}, \mathbf{w}, \boldsymbol{\Xi}$

\STATE \label{prior_f_known} $\boldsymbol{\zeta}_{i}(t-1) = \mathbf{f}(\chol(\hat{\mathbf{\Psi}}(t-1)) \boldsymbol{\xi}_{i} + \hat{\mathbf{x}}(t-1)),$ \\
\hspace*{\fill} $\forall i \in \{1, \cdots, 2M n' \}$

\STATE \label{priorestimateequation} $\mathbf{x}'(t) = \sum_{i = 1}^{2M n'} w_{i} \boldsymbol{\zeta}_{i}(t-1)$

\STATE \label{priorcovequation} $\mathbf{\Psi}'(t) = \sum_{i = 1}^{2M n'} w_{i} \boldsymbol{\zeta}_{i}(t-1) \boldsymbol{\zeta}_{i}^{T}(t-1) - \mathbf{x}'(t) \mathbf{x}'^{T}(t) + \mathbf{\Sigma}$

\ENSURE $\mathbf{x}'(t), \mathbf{\Psi}'(t), \{\boldsymbol{\zeta}_{i}(t-1)| \forall i \in \{1, \cdots, 2M n'\}\}$

\end{algorithmic}
\label{predictionstepCQKF}
\end{algorithm}

\subsubsection{Prediction step}
The \textit{prediction step} computes the prior estimates, ${\mathbf{x}'(t)}$ and ${\mathbf{\Psi}'(t)}$.
At the outset, we compute the Cholesky decomposition \textcolor{black}{\cite{6710599}} of ${\hat{\mathbf{\Psi}}(t-1)}$, which is further put into service to determine the sampling points ${\{\boldsymbol{\zeta}_{i}(t-1)| \forall i \in \{1, \cdots, 2M n'\}\}}$ in step~\ref{prior_f_known}.
At the end, we compute ${\mathbf{x}'(t)}$ and ${\mathbf{\Psi}'(t)}$ by utilizing the updated sampling points in step~\ref{priorestimateequation} and \ref{priorcovequation} of Algorithm~\ref{predictionstepCQKF}.
Here, $\mathbf{x}'(t)$ is an estimated \textbf{mean} vector \cite{9768131}.

Note that CQKF necessitates ${p \in \{0, 1, \cdots, N \}}$ and a random variable (RV) ${\theta \in \mathcal{U}(0,1)}$. The term $p$ is defined in Section~\ref{GoS_subsection}.
If ${p>0}$ and ${\theta  \geq \epsilon'_{p}}$, where ${\epsilon'_{p} = 0.02 \ceil{\frac{p-1}{10}}}$ \cite{9768131}, we advance to the \textit{update step} to compute the posterior estimates, $\hat{\mathbf{x}}(t)$ and $\hat{\mathbf{\Psi}}(t)$.
Otherwise, ${\{ \hat{\mathbf{x}}(t), \hat{\mathbf{\Psi}}(t) \} = \{ \mathbf{x}'(t), \mathbf{\Psi}'(t) \}}$.

\subsubsection{Update step}
In the \textit{update step}, we compute the Cholesky decomposition of ${\mathbf{\Psi}'(t)}$.
Following this, we again determine the sampling points ${\{\boldsymbol{\zeta}_{i}(t)| \forall i \in \{1, \cdots, 2M n'\}\}}$ in step~\ref{posterior_g_known} of Algorithm~\ref{updatestepCQKF}.
Subsequently, we compute a vector ${\hat{\mathbf{y}}(t)}$, representing the predicted sensor measurements, which is then put into service to determine the innovation error covariance ${\mathbf{\Psi}^{*}(t)}$, cross-covariance ${\mathbf{\Psi}^{\dag}(t)}$, and Kalman gain ${\mathbf{K}(t)}$.
Lastly, we compute ${\hat{\mathbf{x}}(t)}$ and ${\hat{\mathbf{\Psi}}(t)}$ by employing ${\mathbf{K}(t)}$, ${\mathbf{\Psi}^{*}(t)}$, and $y(t)$ in step~\ref{posteriorestimateequation} and step~\ref{posteriorcovequation} of Algorithm~\ref{updatestepCQKF}. Here, $y(t)$ and $\hat{\mathbf{x}}(t)$ denote the measurement of the polled sensor and estimated mean vector \cite{9768131}, respectively.

\begin{algorithm}[!t]
\caption{$\updatestep$}\label{updatestepCQKF}
\algsetup{
linenosize=\small,
linenodelimiter=.}
\begin{algorithmic}[1]
\REQUIRE $\mathbf{x}'(t), \mathbf{\Psi}'(t), \mathbf{\Sigma}', \mathbf{w}, \boldsymbol{\Xi}, p, y(t),$ \\
\hspace*{\fill} $\{\boldsymbol{\zeta}_{i}(t-1)| \forall i \in \{1, \cdots, 2M n'\}\}$

\STATE \label{posterior_g_known} $\boldsymbol{\zeta}_{i}(t) = \mathbf{g}(\chol(\mathbf{\Psi}'(t)) \boldsymbol{\xi}_{i} + \mathbf{x}'(t)), \forall i \in \{1, \cdots, 2M n'\}$

\STATE $\hat{\mathbf{y}}(t) = \sum_{i = 1}^{2M n'} w_{i} \boldsymbol{\zeta}_{i}(t)$    \hspace*{\fill} $\triangleright$ $\hat{\mathbf{y}}(t) = [\hat{y}_{1}(t), \cdots, \hat{y}_{N}(t)]^T$

\STATE $\mathbf{\Psi}^{*}(t) = \sum_{i = 1}^{2M n'} w_{i} \boldsymbol{\zeta}_{i}(t) \boldsymbol{\zeta}_{i}^{T}(t) - \hat{\mathbf{y}}(t) \hat{\mathbf{y}}^{T}(t) + \mathbf{\Sigma}'$

\STATE $\mathbf{\Psi}^{\dag}(t) = \sum_{i = 1}^{2M n'} w_{i} \boldsymbol{\zeta}_{i}(t-1) \boldsymbol{\zeta}_{i}^{T}(t) - \mathbf{x}'(t) \hat{\mathbf{y}}^{T}(t)$

\STATE $\mathbf{K}(t) = \mathbf{\Psi}^{\dag}(t) \mathbf{\Psi}^{*}(t)^{-1}$    \hspace*{\fill} $\triangleright$ Kalman gain

\STATE \label{posteriorestimateequation} $\hat{\mathbf{x}}(t) = \mathbf{x}'(t) + \mathbf{K}(t) \mathbf{1}_{p} (y(t)- \hat{y}_{p}(t))$

\STATE \label{posteriorcovequation} $\hat{\mathbf{\Psi}}(t) = \mathbf{\Psi}'(t) - \mathbf{K}(t) \mathbf{\Psi}^{*}(t) \mathbf{K}^{T}(t)$

\ENSURE $\hat{\mathbf{x}}(t), \hat{\mathbf{\Psi}}(t)$

\end{algorithmic}
\label{updatestepCQKF}
\end{algorithm}

\subsection{Query Process and Query Response}\label{sectionqueryprocess}

This work considers MQP.
Each client $c$ operates independently following its own MQP, whose state at time $t$ is denoted as ${q_c(t) \in \mathcal{Q}_c}$.
Here, $\mathcal{Q}_c$ is a set consisting of the MQP's states.
Client $c$ always requests the same query $z_c$ when its corresponding MQP reaches a particular state.
Without loss of generality, we assume the edge node knows the type of query, i.e., $\{z_c, \forall c\}$.
However, the edge node does not know the time steps at which a client would ask a query, instead it observes inter-query durations ${\{\tau'_{c,j} | \forall j \in \{1, \cdots, t'\}, \forall c \in \mathscr{C}\}}$ and the time $\{\tau_{c} \in \mathbb{N} | \forall c \in \mathscr{C}\}$ elapsed since the last query.
Here, $t'$ represents the index for the latest observed inter-query duration instance.
As we know, the clients in $\mathscr{C}$ may not ask queries at the same time step.
Consequently, some elements of the tuple $\{\tau'_{1,t'}, \cdots, \tau'_{C,t'}\}$ may be missing for index $t'$.
To resolve this issue, we have introduced mask $m'_{c,j}$.
If a query is asked by client $c$, then set $m'_{c,t'}=1$, and $\tau'_{c,t'}=\tau_{c}$.
Otherwise, set $m'_{c,t'}=0$ and $\tau'_{c,t'}=\tau^{\dag}$ in the case of missing data, where $\tau^{\dag}$ is a constant.
Subsequently, vector $[\tau'_{1,t'}, m'_{1,t'}, \cdots, \tau'_{C,t'}, m'_{C,t'}]$ is stored in memory buffer $\mathcal{E}'$.
\textcolor{black}{The significance of $m'_{c,t'}$ and $\tau^{\dag}$ is provided later in this section.}

Next, we utilize the LSTM network to estimate tuples $\{(\hat{\tau}_{c, t' +1}, \hat{\sigma}_{c, t' +1}) | \forall c \in \mathscr{C}\}$.
Here, $\hat{\tau}_{c, t' +1}$ and $\hat{\sigma}_{c, t' +1}$ are the estimated inter-query duration and its corresponding estimated standard deviation with respect to the \textit{unobserved} inter-query duration instance $(t' +1)$, respectively.
As stated in step~\ref{step7lstmalgo} of the testing phase in Algorithm~\ref{LSTMalgorithm}, the input to the LSTM network is a sequence
\begin{align}
    & \sequence_{t'} = \{[\tau'_{1,t' - l' + 1}, m'_{1,t' - l' + 1}, \cdots, \tau'_{C,t' - l' + 1}, m'_{C,t' - l' + 1}], \nonumber \\
    & \qquad\qquad\ \cdots \ [\tau'_{1,t'}, m'_{1,t'}, \cdots, \tau'_{C,t'}, m'_{C,t'}]\},
\end{align}
while its output is ${\{(\hat{\tau}_{c, t' +1}, \hat{\sigma}_{c, t' +1}) | \forall c \in \mathscr{C}\}}$.
Here $l'$ is the sequence length.
If a query is asked by client $c$, then retain the recently estimated tuple $(\hat{\tau}_{c, t' +1}, \hat{\sigma}_{c, t' +1})$; otherwise, $(\hat{\tau}_{c, t' +1}, \hat{\sigma}_{c, t' +1}) = (\hat{\tau}_{c, t'}, \hat{\sigma}_{c, t'})$.

\begin{algorithm}[!t]
\caption{LSTM}\label{LSTMalgorithm}
\algsetup{
linenosize=\small,
linenodelimiter=.}
\begin{algorithmic}[1]
\REQUIRE $t', \{(\hat{\tau}_{c, t'}, \hat{\sigma}_{c, t'}) | \forall c \in \mathscr{C}\}$

\TRAINING
\FOR{$\floor{(|\mathcal{E}'| - l')/|\mathcal{B}'|}$}
\STATE Sample a mini-batch $\mathcal{B}'$ of sequences ${\{\sequence_{j} | \forall j \in \{t'-|\mathcal{B}'|+1, \cdots, t'\}\}}$ from $\mathcal{E}'$.
\STATE Provide $\mathcal{B}'$ as input to the LSTM network and update the LSTM network by minimizing $\Omega'$, in \textcolor{black}{$(\ref{LSTMlossequation_final})$,} using the Adam optimizer
\ENDFOR
\end{algorithmic}

\begin{algorithmic}[1]
\TESTING
\IF{query asked by at least one client}
\STATE Sample a sequence $\sequence_{t'}$ from $\mathcal{E}'$.
\STATE \label{step7lstmalgo} Provide $\sequence_{t'}$ as input to the LSTM network and obtain ${\{(\hat{\tau}_{c, t' +1}, \hat{\sigma}_{c, t' +1}) | \forall c \in \mathscr{C}\}}$ as output
\FOR{$c \in \mathscr{C}$}
    \IF{query asked by client $c$}
    \STATE Retain recently estimated $(\hat{\tau}_{c, t' +1}, \hat{\sigma}_{c, t' +1})$
    \ELSE
    \STATE $(\hat{\tau}_{c, t' +1}, \hat{\sigma}_{c, t' +1}) = (\hat{\tau}_{c, t'}, \hat{\sigma}_{c, t'})$
    \ENDIF
\ENDFOR
\ELSE
\STATE $(\hat{\tau}_{c, t' +1}, \hat{\sigma}_{c, t' +1}) = (\hat{\tau}_{c, t'}, \hat{\sigma}_{c, t'}), \forall c \in \mathscr{C}$
\ENDIF
\ENSURE $\{(\hat{\tau}_{c, t' +1}, \hat{\sigma}_{c, t' +1}) | \forall c \in \mathscr{C}\}$

\end{algorithmic}
\label{LSTMalgorithm}
\end{algorithm}

As described in Algorithm~\ref{LSTMalgorithm}, the training process for the LSTM network commences by sampling a mini-batch $\mathcal{B}'$ of sequences ${\{\sequence_{j} | \forall j \in \{t'-|\mathcal{B}'|+1, \cdots, t'\}\}}$ from $\mathcal{E}'$.
Then, we provide $\mathcal{B}'$ as input to the LSTM network and obtain output ${\{(\hat{\tau}_{c, t' +1,j}, \hat{\sigma}_{c, t' +1,j}) | \forall j \in \{t'-|\mathcal{B}'|+1, \cdots, t'\}, \forall c\}}$.
Thereupon, we update the LSTM network by minimizing the Gaussian negative log-likelihood loss \cite{annurev_statistics_042424_050626, NIPS2017_2650d608} $(\Omega')$ using the adaptive moment estimation (Adam) optimizer \cite{kingma2014adam}, where
\begin{subequations}
\begin{align}
    \label{LSTMlossequation}\Omega'_{c,j} =& \ \frac{1}{2} \Bigl( \log(2\pi \hat{\sigma}_{c, t' +1,j}^{2}) + \frac{(\tau'_{c, t' +1} - \hat{\tau}_{c, t' +1,j})^2}{\hat{\sigma}_{c, t' +1,j}^{2}} \Bigr) , \\
    \label{LSTMlossequation_final}\Omega' =& \ \frac{1}{|\mathcal{B}'|} \sum_{j=t'-|\mathcal{B}'|+1}^{t'} \sum_{c=1}^{C} m'_{c,t'+1} \Omega'_{c,j} .
\end{align}
\end{subequations}
This training process is repeated $\floor{(|\mathcal{E}'| - l')/|\mathcal{B}'|}$ times.
Note that feeding $m'_{c,t'}$ allows the LSTM network to learn to ignore the specific input feature, i.e., $\tau'_{c,t'}$, when $m'_{c,t'}=0$ because during training $\Omega'_{c,j}$ never contributes to the loss when $m'_{c,t'+1}=0$ in $(\ref{LSTMlossequation_final})$.
Meanwhile, setting $\tau'_{c,t'}=\tau^{\dag}$ lets the LSTM network run even in the case of missing data.

The edge node responds to a query, from client ${c \in \mathscr{C}}$, with an estimate ${\hat{z}_{c}(\hat{\mathbf{x}}(t), \hat{\mathbf{\Psi}}(t))}$.
The objective of the edge node is to respond to queries as accurately as possible, essentially minimizing the error in query responses.
This error is quantified by the $\MSE$ of the query response, which for client $c$ is defined as \cite{9768131, 10143239}
\begin{align}\label{MSE_queryPosterior}
    \widehat{\MSE}_{c}(t) = \mathbb{E} \bigl[(\hat{z}_{c}(\hat{\mathbf{x}}(t), \hat{\mathbf{\Psi}}(t)) - z_{c}(\mathbf{x}(t)))^2 \bigr].
\end{align}

\subsection{GoS Problem}\label{GoS_subsection}

The problem is to anticipate future queries and schedule sensor transmissions to minimize the $\MSE$ on future query responses.
This task demands foresight, necessitating an understanding not only of the monitored NLDS and NLOM but also of the query process and the interplay among various query functions.

We can model the GoS problem at the edge node as a partially observable Markov decision process (POMDP), in which the edge node must decide whether to poll a sensor.
Herein, the \textit{action space} is ${\mathcal{A} = \{0, 1, \cdots, N \}}$, where action ${p=0}$ signifies no sensor is polled, and action ${p=n \in \{ 1, \cdots , N\}}$ represents sensor $n$ is polled.

Before initiating the sensor scheduling operation, the edge node possesses prior estimates ${\{\mathbf{x}'(t), \mathbf{\Psi}'(t)\}}$.
In \cite{10143239}, the authors incorporated these prior estimates as a component of the POMDP's state.
Here, ${\mathbf{x}'(t)}$ and ${\mathbf{\Psi}'(t)}$ are the expected value and covariance matrix of the prior state estimate ${\mathcal{N}(\mathbf{x}'(t), \mathbf{\Psi}'(t))}$, respectively.
Note that CQKF is a recursive algorithm, and it does not store the past state estimates/measurements explicitly.
Instead, at the prediction/update step, CQKF folds the influence of the past state estimates/measurements into the mean vector and covariance matrix.
Thus, ${\mathbf{x}'(t)}$ encapsulates the aforementioned influence in the prior state estimate ${\mathcal{N}(\mathbf{x}'(t), \mathbf{\Psi}'(t))}$.
Likewise, ${\mathbf{\Psi}'(t)}$ encapsulates the aforementioned influence onto the second-order uncertainty in the prior state estimate ${\mathcal{N}(\mathbf{x}'(t), \mathbf{\Psi}'(t))}$.

As shown \textcolor{black}{in Appendix}:
(i) the strongest eigenvector $\boldsymbol{e}$ of ${\mathbf{\Psi}'(t)}$ indicates the principal direction in the prior estimated state space along which the uncertainty is maximum, and
(ii) the strongest eigenvalue $\lambda$ of ${\mathbf{\Psi}'(t)}$ is the aforesaid uncertainty along $\boldsymbol{e}$.
In addition, \textcolor{black}{the} elements of the product ${\lambda \boldsymbol{e}}$ represent the contribution of the different components of the prior state estimate to the uncertainty \textcolor{black}{along direction} $\boldsymbol{e}$.

The goal is to minimize the $\MSE$ of the query \textcolor{black}{response;} thus, it is reasonable to incorporate
\begin{align}\label{MSEpriorEstimates}
    \MSE'_{c}(t) = \mathbb{E} \bigl[(\hat{z}_{c}(\mathbf{x}'(t), \mathbf{\Psi}'(t)) - z_{c}(\mathbf{x}(t)))^2 \bigr],
\end{align}
\textcolor{black}{${\forall c \in \mathscr{C}}$, in POMDP's state to inform} the scheduler about the
(i) $\MSE$ of the query response with respect to the existing prior estimates, and
(ii) out-turn of adopting \textcolor{black}{${\textrm{action 0}}$} on the $\MSE$ of the query response.
However, incorporating ${\MSE'_{c}(t)}$ in POMDP's state would require ${\MSE'_{c}(t)}$ to be computed even at timesteps with no queries.

We represent POMDP's \textit{state} in the following two ways, ${\boldsymbol{s}'(t) = (\mathbf{x}'(t), \mathbf{\Psi}'(t), \{\boldsymbol{d}_{c} | \forall c \in \mathscr{C}\})}$ and ${\boldsymbol{s}^{*}(t) = (\mathbf{x}'(t), \lambda \boldsymbol{e}, \{\MSE'_{c}(t), \boldsymbol{d}_{c} | \forall c \in \mathscr{C}\})}$.
If client $c$ asks a query, then $\boldsymbol{d}_{c} = [0, 0, 0]^{T}$; otherwise, $\boldsymbol{d}_{c} = [\tau_{c}, \hat{\tau}_{c, t' +1}, \hat{\sigma}_{c, t' +1}]^{T}$, \textcolor{black}{where $\{\hat{\tau}_{c, t' +1}, \hat{\sigma}_{c, t' +1}\}$ are computed using the trained LSTM network.}
The state space \textcolor{black}{$\mathcal{S}$} of ${\boldsymbol{s}'(t)}$ and ${\boldsymbol{s}^{*}(t)}$ is ${\mathbb{R}^{M^{2}+M} \times \mathbb{R}_{+}^{C} \times \mathbb{R}_{\geq0}^{C} \times \mathbb{N}^{C}}$ and ${\mathbb{R}^{2M+C} \times \mathbb{R}_{+}^{C} \times \mathbb{R}_{\geq0}^{C} \times \mathbb{N}^{C}}$, respectively.
Note that $\MSE'_{c}(t)$, $\forall c \in \mathscr{C}$, is a deterministic function of ${\{\mathbf{x}'(t), \mathbf{\Psi}'(t)\}}$.
Because of this, $\MSE'_{c}(t)$, $\forall c \in \mathscr{C}$, is not a part of $\boldsymbol{s}'(t)$.
\textcolor{black}{Henceforth}, discussions are valid for both ${\boldsymbol{s}'(t)}$ and ${\boldsymbol{s}^{*}(t)}$, hence we use the subscript-less notation ${\boldsymbol{s}(t)}$.

The \textit{reward} ${r(t)}$ in POMDP is defined as
\begin{align}\label{rewardequation}
   r' = & - \sum_{c=1}^{C} d^{\mathbbm{1}(\tau_{c}=0)} e^{-d' \hat{\tau}_{c, t' +1}} \widehat{\MSE}_{c}(t) - d^{*}\hat{\sigma}_{c, t' +1} , \nonumber \\
   r^{*} = & \sum_{c=1}^{C} d^{\mathbbm{1}(\tau_{c}=0)} e^{-d' \hat{\tau}_{c, t' +1}} \bigl(\MSE'_{c}(t) \nonumber \\
   & \qquad \qquad \qquad \qquad \qquad - \widehat{\MSE}_{c}(t)\bigr) - d^{*}\hat{\sigma}_{c, t' +1} , \nonumber \\
   r^{\dag} = & \ r^{*} - d^{\dag} , \nonumber \\
   r(t) = & \mathbbm{1}(p=0) r' + \mathbbm{1}(p \mathrel{\ne} 0) r^{\dag} ,
\end{align}
where ${p \in \mathcal{A}}$ denotes the selected \textit{action}, \textcolor{black}{and} $d^{\dag}$ denotes transmission cost.
Note that $(\ref{rewardequation})$ demands ${\widehat{\MSE}_{c}(t)}$ to be computed even at timesteps with no queries.
In ${r(t)}$, the term ${(\MSE'_{c}(t) - \widehat{\MSE}_{c}(t))}$ represents the $\voi$ of the fetched data \cite{10273599, li2023towards}.
Next, the term $e^{- d' \hat{\tau}_{c, t' +1}}$ incentivizes the scheduler to reduce the $\MSE$ of the query response when queries are imminent, while the term $d^{*} \hat{\sigma}_{c, t' +1}$ is a penalty for uncertainty in the inter-query duration estimates.
\textcolor{black}{We assume ${0<d^{*}\ll1}$ and ${0<d'<1}$ to make ${\widehat{\MSE}_{c}(t)}$ and ${(\MSE'_{c}(t) - \widehat{\MSE}_{c}(t))}$ relatively more significant than $\hat{\sigma}_{c, t' +1}$.}
This approach allows the scheduler to get prepared for timesteps with queries in a proactive manner.
Besides, for ${d > 1}$, the term $d^{\mathbbm{1}(\tau_{c}=0)}$ informs the scheduler that timesteps with queries are of relatively higher importance.
Furthermore, to incentivize the scheduler to opt for \textcolor{black}{${\textrm{action 0}}$} when $d^{\dag}$ becomes significantly higher than $r^{*}$, term ${\mathbbm{1}(p=0) r'}$ is incorporated in ${r(t)}$.

The long-term reward $R(\pi)$ can be stated as
\begin{align}\label{longTerm_reward_eq}
    R(\pi(t)) = \mathbb{E} \Biggl[ \sum_{t^{\dag}=0}^{\infty} \gamma^{t^{\dag}} r(t + t^{\dag}) \bigg| \boldsymbol{s}(\textcolor{black}{t}) , \pi(t) \Biggr],
\end{align}
where ${\gamma \in [0,1)}$ is the exponential discount factor. Moreover, ${\pi : \textcolor{black}{\mathcal{S}} \rightarrow \mathbf{\Phi}(\mathcal{A})}$ represents the policy which maps \textcolor{black}{$\mathcal{S}$} to ${\mathbf{\Phi}(\mathcal{A})}$, where ${\mathbf{\Phi}(\mathcal{A})}$ \textcolor{black}{encompasses the probability of selecting each action.}
Finally, the GoS problem can be defined as \cite{10143239}
\begin{align}\label{schedulingProblem}
    \pi^{*}(t) = \underset{\pi : \textcolor{black}{\mathcal{S}} \rightarrow \mathbf{\Phi}(\mathcal{A})}{\argmax} R(\pi(t)),
\end{align}
where $\pi^{*}$ represents the optimal policy.

\subsection{CQKF-cum-DRL-based Scheduler}\label{subsection_CQKF_cum_DRL_based_Scheduler}

\begin{figure}[!t]
\centering
\includegraphics[width=\linewidth]{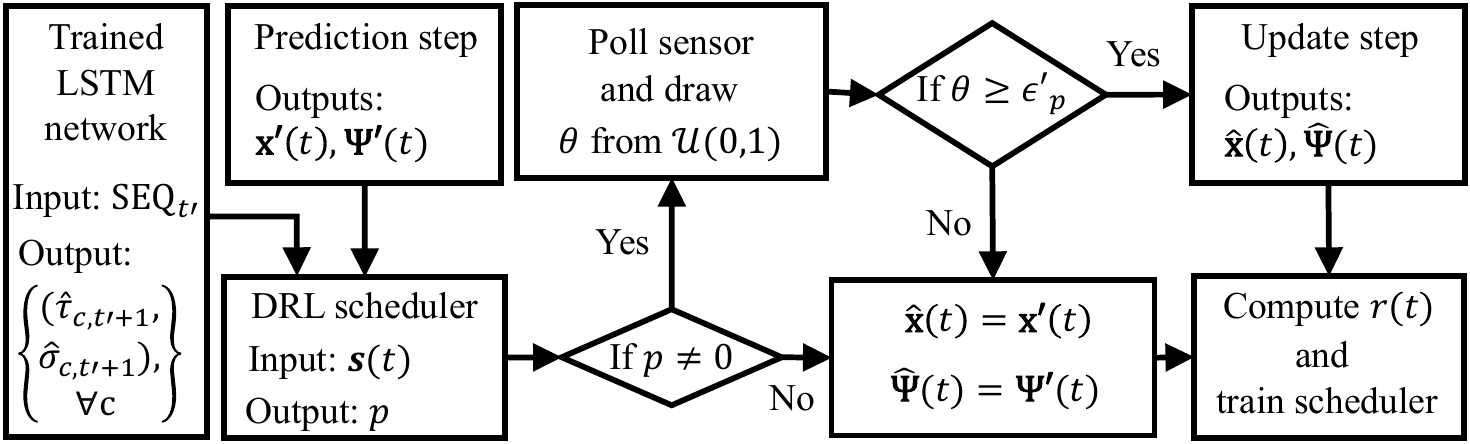}
\caption{\textcolor{black}{A schematic of our proposed GoS.}}
\label{schematicfigure}
\end{figure}

We solve $(\ref{schedulingProblem})$ using \textcolor{black}{DRL;} thus, we name our scheduler as CQKF-cum-DRL-based scheduler, described in detail in Algorithm~\ref{CQKF_DRLscheduler}.
Meanwhile, we are maintaining two DNNs, named the online network and the target network, to improve the stability of the proposed DRL scheduler.
A schematic of the proposed GoS is shown in Fig.~\ref{schematicfigure}.

Algorithm~\ref{CQKF_DRLscheduler} operates as follows.
Initially, it computes $\{(\tau_{c}, \hat{\tau}_{c, t' +1}, \hat{\sigma}_{c, t' +1}, \MSE'_{c}(t), \boldsymbol{d}_{c}) | \forall c \in \mathscr{C}\}$ and prior estimates to formulate ${\boldsymbol{s}(t)}$.
\textcolor{black}{Here, $\{\hat{\tau}_{c, t' +1}, \hat{\sigma}_{c, t' +1}\}$ are computed using the trained LSTM network, while the} computation of $\MSE'_{c}(t)$ involves taking $S$ samples from a Gaussian distribution with mean $\mathbf{x}'(t)$ and covariance $\mathbf{\Psi}'(t)$.
These samples are then utilized to obtain the vector $\boldsymbol{u} = [ u_{1}, \cdots, u_{S} ]^T$, where $u_s = z_{c}(\mathbf{x}_s)$ and $\mathbf{x}_s$ is the $s^{th}$ sample.
The variance of $\boldsymbol{u}$ yields $\MSE'_{c}(t)$.
Subsequently, the online network, characterized by its weights \textcolor{black}{$\Theta_{1}$}, takes ${\boldsymbol{s}(t)}$ as its input and outputs the \textit{action values} ${\hat{q}_{i}(\boldsymbol{s}(t)),} {\forall i \in \mathcal{A}}$.
Here, ${\hat{q}_{i}(\boldsymbol{s}(t))}$ serves as an estimate of the reward that the scheduler would gain if action $i$ is chosen \cite{raghuwanshi2024neural}.
The $\epsilon$-greedy method \cite{raghuwanshi2024neural} then employs the action values to select an action ${p \in \mathcal{A}}$.
Primarily, the $\epsilon$-greedy method opts to select $p$ as the argument of the maximum action value. However, to explore the whole action space, the $\epsilon$-greedy method occasionally opts to select $p$ randomly from the set $\mathcal{A}$.
The former operation is called exploitation, while the \textcolor{black}{latter is} exploration.
The posterior estimates are then reckoned according to steps~\ref{step2cqkfalgo}-\ref{step7cqkfalgo} of Algorithm~\ref{CQKFalgorithm}.
Subsequently, compute ${\widehat{\MSE}_{c}(t)}$, ${\forall c \in \mathscr{C}}$, with respect to the procedure available in Algorithm~\ref{rewardalgo}, and reckon $r(t)$ using $(\ref{rewardequation})$.
Here, ${r(t)}$ is the reward gained by the online network for selecting action $p$.

\begin{algorithm*}[!t]
\caption{CQKF-cum-DRL-based Scheduler at $t$}\label{CQKF_DRLscheduler}
\algsetup{
linenosize=\small,
linenodelimiter=.}
\begin{multicols}{2}
\begin{algorithmic}[1]

\REQUIRE $\textcolor{black}{\Theta_{2}}, \boldsymbol{s}(t-1), \hat{\mathbf{x}}(t-1), \hat{\mathbf{\Psi}}(t-1), \eta, \eta', \eta^{*}, \epsilon,$\\
\hspace*{\fill}  $t', \{(\tau_{c}, \hat{\tau}_{c, t'}, \hat{\sigma}_{c, t'}) | \forall c \in \mathscr{C}\}$

\STATE Compute $\{(\hat{\tau}_{c, t' +1}, \hat{\sigma}_{c, t' +1}) | \forall c \in \mathscr{C}\}$ using Algorithm~\ref{LSTMalgorithm}
\STATE Compute $\{\mathbf{x}'(t), \mathbf{\Psi}'(t), \{\boldsymbol{\zeta}_{i}(t-1)| \forall i \in \{1, \cdots, 2M n'\}\}\}$ using step~\ref{step1cqkfalgo} of Algorithm~\ref{CQKFalgorithm}
\FOR{$c \in \mathscr{C}$}
    \IF{query asked by client $c$}
    \STATE $\boldsymbol{d}_{c} = [0, 0, 0]^{T}$
    \ELSE
    \STATE $\boldsymbol{d}_{c} = [\tau_{c}, \hat{\tau}_{c, t' +1}, \hat{\sigma}_{c, t' +1}]^{T}$
    \ENDIF
    \STATE Draw $\mathbf{x}_s$ from ${\mathcal{N}(\mathbf{x}'(t), \mathbf{\Psi}'(t)), \forall s \in \{1, \cdots, S \}}$
    \STATE $u_s = z_{c}(\mathbf{x}_s), \forall s \in \{1, \cdots, S \}$    \hspace*{\fill} $\triangleright$ $\boldsymbol{u} = [ u_{1}, \cdots, u_{S} ]^T$
    \STATE $\MSE'_{c}(t) = \VAR(\boldsymbol{u})$ \hspace*{\fill} $\triangleright$ Sample variance
\ENDFOR

\STATE Evaluate $\hat{q}_{i}(\boldsymbol{s}(t)), \forall i \in \mathcal{A}$ using the online network

\STATE Draw $\theta$ from $\mathcal{U}(0,1)$

\IF{$\theta > \epsilon$} \label{step4proposedscheduleralgo}
  \STATE $p \leftarrow \argmax_{i\in \mathcal{A}} \hat{q}_{i}(\boldsymbol{s}(t))$    \hspace*{\fill} $\triangleright$ Exploitation
\ELSE
  \STATE \label{step7proposedscheduleralgo} Select $p$ randomly from $\{ 0, \cdots, N \}$    \hspace*{\fill} $\triangleright$ Exploration
\ENDIF    \hspace*{\fill} $\triangleright$ $p:$ index of selected action

\STATE Compute ${\{\hat{\mathbf{x}}(t), \hat{\mathbf{\Psi}}(t)\}}$ using steps~\ref{step2cqkfalgo}-\ref{step7cqkfalgo} of Algorithm~\ref{CQKFalgorithm}

\STATE $r(t) \leftarrow \reward(\mathscr{C}, S, \hat{\mathbf{x}}(t), \hat{\mathbf{\Psi}}(t), p, d^{\dag},$ \\
\hspace*{\fill} ${\{\tau_{c}, \MSE'_{c}(t) | \forall c \}})$

\IF{$\eta = |\mathcal{E}|$}
  \STATE Remove $T_{|\mathcal{B}|}$ from $\mathcal{E}$
  \STATE $\eta \leftarrow \eta - 1$
\ENDIF    \hspace*{\fill} $\triangleright$ $\mathcal{E}:$ memory buffer at the edge node

\IF{$t>1$}
  \STATE \label{tuplestore} Store $\{ \boldsymbol{s}(t-1), p, r(t), \boldsymbol{s}(t) \}$ as ${(\eta + 1)^{th}}$ tuple in $\mathcal{E}$

  \STATE $\eta \leftarrow \eta + 1$    \hspace*{\fill} $\triangleright$ $\eta:$ number of tuples available in $\mathcal{E}$
\ENDIF

\STATE $\eta' \leftarrow \eta' + 1$
\IF{$\eta' = \eta'_{max}$}
  \STATE $\textcolor{black}{\Theta_{2} = \Theta_{1}}$  \hspace*{\fill} $\triangleright$ Update target network
  \STATE $\eta' = 0$  \hspace*{\fill} $\triangleright$ Restart counter
\ENDIF

\STATE Sample a mini-batch $\mathcal{B}$ of size $|\mathcal{B}|$ from $\mathcal{E}$. Then, provide ${T_{j,4}, \forall j \in \{1, \cdots , |\mathcal{B}|\}}$, as input to the target network and utilize the target network's outputs in $(\ref{targetvalueequation})$ to determine the target values ${\Bar{\Bar{q}}_{j}, \forall j \in \{1, \cdots , |\mathcal{B}|\}}$ for $\mathcal{B}$

\STATE Provide ${T_{j,1}, \forall j \in \{1, \cdots , |\mathcal{B}|\}}$, as input to the online network and utilize the corresponding target values ${\Bar{\Bar{q}}_{j}, \forall j \in \{1, \cdots , |\mathcal{B}|\}}$, as labels for updating \textcolor{black}{$\Theta_{1}$} by minimizing $\Omega$, in $(\ref{DRLlossequation})$, using RMSProp

\IF{$t > 2|\mathcal{B}|$}\label{step37_algo6}
\IF{$\eta^{*} = \eta^{*}_{max}$}
  \STATE $\epsilon \leftarrow \max(0.15, \ \epsilon - 0.05)$
  \STATE $\eta^{*} = 0$  \hspace*{\fill} $\triangleright$ Restart counter
\ENDIF
\STATE $\eta^{*} \leftarrow \eta^{*} + 1$
\ENDIF\label{step43_algo6}

\ENSURE $\hat{\mathbf{x}}(t), \hat{\mathbf{\Psi}}(t), \epsilon, \textcolor{black}{\Theta_{2}}, \boldsymbol{s}(t), \eta, \eta', \eta^{*}, t',$ \\
\hspace*{\fill}  $\{(\tau_{c}, \hat{\tau}_{c, t' +1}, \hat{\sigma}_{c, t' +1}) | \forall c \in \mathscr{C}\}$

\end{algorithmic}
\end{multicols}
\label{CQKF_DRLscheduler}
\end{algorithm*}

Now that both $r(t)$ and $p$ are available, we proceed to store ${\{ \boldsymbol{s}(t-1), p, r(t), \boldsymbol{s}(t) \}}$ as $T_{\eta + 1}$, i.e., ${(\eta + 1)^{th}}$ tuple, in the finite memory $\mathcal{E}$ and increase $\eta$ by $1$. Here, $\eta$ represents the number of tuples available in $\mathcal{E}$.
If $\mathcal{E}$ is full, we remove $T_{|\mathcal{B}|}$ from $\mathcal{E}$ and decrease $\eta$ by $1$ before storing the new tuple.
Following this, we update the target network weights, denoted as \textcolor{black}{$\Theta_{2}$}, by setting \textcolor{black}{${\Theta_{2} = \Theta_{1}}$}, if the counter $\eta'$ reached its threshold value, herein set to $\eta'_{max}$.

Next, the training process for the online network commences by sampling a mini-batch $\mathcal{B}$ of size $|\mathcal{B}|$ from $\mathcal{E}$.
Then, we provide ${T_{j,4}, \forall j \in \{1, \cdots , |\mathcal{B}|\}}$, i.e., \textcolor{black}{the} fourth element of ${T_{j} \in \mathcal{B}}$, as input to the target network and obtain its output ${\vec{\boldsymbol{q}}_{j} = \{\vec{q}_{j,i} | \forall i \in \mathcal{A}\},} {\forall j \in \{1, \cdots , |\mathcal{B}|\}}$.
Now, we utilize outputs of the target network to determine the target values as
\begin{align}\label{targetvalueequation}
    \Bar{\Bar{q}}_{j} = T_{j,3} + \gamma \underset{i \in \mathcal{A}}{\max} \ \vec{q}_{j,i} , \forall j \in \{1, \cdots , \textcolor{black}{|\mathcal{B}|}\}, 
\end{align}
for $\mathcal{B}$.
\textcolor{black}{In addition}, ${\Bar{\Bar{q}}_{j}, \forall j \in \{1, \cdots , |\mathcal{B}|\}}$, is an estimate of $R(\pi)$.
Thereupon, we provide ${T_{j,1}, \forall j \in \{1, \cdots , |\mathcal{B}|\}}$, as input to the online network.
The corresponding target values ${\Bar{\Bar{q}}_{j}, \forall j \in \{1, \cdots , \textcolor{black}{|\mathcal{B}|}\}}$, serve as labels for updating \textcolor{black}{$\Theta_{1}$} by minimizing $\Omega$ using RMSProp \cite{raghuwanshi2024neural} optimizer, where
\begin{align}\label{DRLlossequation}
    \Omega = \frac{1}{|\mathcal{B}|} \sum_{j=1}^{|\mathcal{B}|} \big(\Bar{\Bar{q}}_{j} - \hat{q}_{T_{j,2}}(T_{j,1}) \big)^2.
\end{align}

To deal with the exploding gradient problem during the online network's training phase, we perform the gradient-norm clipping \textcolor{black}{\cite{gradient_clipping}}. This involves clipping the gradient vector $\nabla_{\textcolor{black}{\Theta_{1}}} \Omega$ as
\begin{equation}\label{eq5}
    \boldsymbol{\chi} = \frac{a' \ \nabla_{\textcolor{black}{\Theta_{1}}} \Omega}{\max({\Vert \nabla_{\textcolor{black}{\Theta_{1}}} \Omega \Vert}_2, a')}  .
\end{equation}
Herein, $a'$ represents the threshold value for ${\Vert \nabla_{\textcolor{black}{\Theta_{1}}} \Omega \Vert}_2$ and the vector $\boldsymbol{\chi}$ stores the clipped gradients.
At last, to emphasize exploitation over exploration in the $\epsilon$-greedy method, it is necessary to gradually decrease $\epsilon$. Thus, we reduce $\epsilon$ by $0.05$, unless it has already reached $0.15$.

\begin{algorithm}[!t]
\caption{$\reward$}\label{rewardalgo}
\algsetup{
linenosize=\small,
linenodelimiter=.}
\begin{algorithmic}[1]
\REQUIRE ${\mathscr{C}, S, \hat{\mathbf{x}}(t), \hat{\mathbf{\Psi}}(t), p, d^{\dag}, \{\tau_{c}, \MSE'_{c}(t) | \forall c \}}$

\IF{$p=0$}
\STATE \label{step2rewardalgo} $\widehat{\MSE}_{c}(t) = \MSE'_{c}(t), \forall c \in \mathscr{C}$
\ELSE
\FOR{$c \in \mathscr{C}$}
   \STATE \label{step4rewardalgo} Draw $\mathbf{x}_s$ from ${\mathcal{N}(\hat{\mathbf{x}}(t), \hat{\mathbf{\Psi}}(t)), \forall s \in \{1, \cdots, S \}}$
   
   \STATE \label{step7rewardalgo} $u_s = z_{c}(\mathbf{x}_s), \forall s \in \{1, \cdots, S \}$    \hspace*{\fill} $\triangleright$ $\boldsymbol{u} = [ u_{1}, \cdots, u_{S} ]^T$

  \STATE \label{step8rewardalgo} $\widehat{\MSE}_{c}(t) = \VAR(\boldsymbol{u})$ \hspace*{\fill} $\triangleright$ Sample variance
\ENDFOR
\ENDIF

\STATE Compute ${r(t)}$ using $(\ref{rewardequation})$  \hspace*{\fill} $\triangleright$ Reward

\ENSURE $r(t)$

\end{algorithmic}
\label{rewardalgo}
\end{algorithm}

\section{Benchmark Schedulers} \label{benchmarkschedulersection}

\subsection{Monte Carlo scheduler} \label{montecarlosubsection}

The Monte Carlo scheduler, described in detail in Algorithm~\ref{montecarloscheduler}, is adopted as a benchmark due to its versatility in handling any query type.
For a given client ${c \in \mathscr{C}}$, Algorithm~\ref{montecarloscheduler} initially computes the prior estimates, and then subsequently, in an iterative manner, $S$ distinct Gaussian samples are drawn for sensor $n$ in step~\ref{step11montecarloalgo}, by computing $S$ distinct posterior estimates either in step~\ref{step7montecarloalgo} or in step~\ref{step9montecarloalgo} depending on the inequality in step~\ref{step5MonteCarloalgo}.
The $S$ Gaussian samples are then employed to compute $S$ distinct query responses in step~\ref{step12montecarloalgo}, in an iterative manner.
These query responses are stored in ${\boldsymbol{u}}$.
Next, in step~\ref{step14montecarloalgo}, ${\VAR(\boldsymbol{u})}$ is computed and stored in ${\boldsymbol{\nu}}$. Here ${\VAR(\boldsymbol{u})}$ represents ${\widehat{\MSE}_{c}(t)}$ expected in case sensor $n$ is polled.
Repeat the procedure outlined from step~\ref{step3montecarloalgo} to step~\ref{step14montecarloalgo} a total of $N$ times, to calculate ${\VAR(\boldsymbol{u})}$ for every sensor.
Now, in step~\ref{step16montecarloalgo}, a sensor is polled, whose index value corresponds to the index of the minimum element in ${\boldsymbol{\nu}}$.
Following this, to compute the actual ${\widehat{\MSE}_{c}(t)}$ in step~\ref{step18montecarloalgo}, Algorithm~\ref{montecarloscheduler} again computes the posterior estimates by leveraging the received observation from the polled sensor.

Indeed, it is worth mentioning that the Monte Carlo scheduler does come with limitations. Unlike the proposed CQKF-cum-DRL-based scheduler, we need to design $C$ Monte Carlo schedulers in the case of $C$ clients.
Moreover, the Monte Carlo scheduler does not even take into account the information related to the query requests while polling a sensor.
It simply polls a sensor whenever a query is asked.

\begin{algorithm}[!t]
\caption{Monte Carlo Scheduler for Client $c \in \mathscr{C}$}\label{montecarloscheduler}
\algsetup{
linenosize=\small,
linenodelimiter=.}
\begin{algorithmic}[1]

\REQUIRE $\hat{\mathbf{x}} = \hat{\mathbf{x}}(t-1), \hat{\mathbf{\Psi}} = \hat{\mathbf{\Psi}}(t-1)$

\FOR{$n \in \{1, \cdots, N \}$} \label{step2montecarloalgo}
    \FOR{$s \in \{1, \cdots, S \}$} \label{step3montecarloalgo}
       \STATE $\mathbf{x}', \mathbf{\Psi}', \mathbf{Z}^{*} \leftarrow \predictionstep(\hat{\mathbf{x}}, \hat{\mathbf{\Psi}}, \mathbf{\Sigma}, \mathbf{w}, \boldsymbol{\Xi})$
        
       \STATE Draw $\theta$ from $\mathcal{U}(0,1)$
       \IF{$\theta  \geq 0.02 \ceil{\frac{n-1}{10}}$} \label{step5MonteCarloalgo}
          \STATE Draw $y$ from $\mathcal{N}(\mathbf{1}_{n}^{T}\hat{\mathbf{x}}, \mathbf{1}_{n}^{T} \hat{\mathbf{\Psi}} \mathbf{1}_{n})$

          \STATE \label{step7montecarloalgo} $\hat{\mathbf{x}}, \hat{\mathbf{\Psi}} \leftarrow \updatestep(\mathbf{x}', \mathbf{\Psi}', \mathbf{Z}^{*}, \mathbf{\Sigma}', \mathbf{w}, \boldsymbol{\Xi}, y, n)$
       \ELSE
            \STATE \label{step9montecarloalgo} $\{\hat{\mathbf{x}}, \hat{\mathbf{\Psi}}\} = \{\mathbf{x}', \mathbf{\Psi}'\}$
       \ENDIF
       \STATE \label{step11montecarloalgo} $\mathbf{x}_s = \mathcal{N}(\hat{\mathbf{x}}, \hat{\mathbf{\Psi}})$
       \STATE \label{step12montecarloalgo} $u_s = z_{c}(\mathbf{x}_s)$
    \ENDFOR
    \STATE \label{step14montecarloalgo} $\nu_n = \VAR(\boldsymbol{u})$ \hspace*{\fill} $\triangleright$ Sample variance
\ENDFOR

\STATE \label{step16montecarloalgo} $p = \argmin_{n \in \{ 1, \cdots, N \}} \boldsymbol{\nu}$    \hspace*{\fill} $\triangleright$ $\boldsymbol{\nu} = [ \nu_{1}, \cdots, \nu_{N} ]^T$

\STATE \label{step17montecarloalgo} Compute ${\{\hat{\mathbf{x}}(t), \hat{\mathbf{\Psi}}(t)\}}$ using steps~\ref{step2cqkfalgo}-\ref{step7cqkfalgo} of Algorithm~\ref{CQKFalgorithm}

\STATE \label{step18montecarloalgo} Compute ${\widehat{\MSE}_{c}(t)}$ using steps~\ref{step4rewardalgo}-\ref{step8rewardalgo} of Algorithm~\ref{rewardalgo}

\ENSURE $\hat{\mathbf{x}}(t), \hat{\mathbf{\Psi}}(t)$

\end{algorithmic}
\label{montecarloscheduler}
\end{algorithm}

\subsection{Benchmark DRL Scheduler} \label{benchmarksche2subsection}

Our second benchmark scheduler adopts the action space, POMDP state/observation space, reward function, and online and target network architecture utilized by the scheduler in \cite{10143239}.
The working of the benchmark DRL scheduler is same as the one described in Algorithm~\ref{CQKF_DRLscheduler}, except for the following changes:
\begin{itemize}
\item Change $\mathcal{A}$ to ${\{1, \cdots, N \}}$, indicating that the edge node must poll a sensor at every time step.
\item In Algorithm~\ref{CQKF_DRLscheduler}, provide ${\boldsymbol{s}(t) = (\mathbf{x}'(t), \mathbf{\Psi}'(t), \{\tau_{c}, \forall c\})}$, with ${\mathcal{S} = \mathbb{R}^{M + M^2} \times \mathbb{N}^C}$, as input to the online network.
\item When no query is posed at $t$, reward ${r(t) = 0}$.
\item Change the online and target network architecture by increasing the number of hidden layers to three, having ${\{ 2.5M , M , N\}}$ neurons and a dropout probability of ${\{ 0.1 , 0.1 , 0\}}$, respectively.
\end{itemize}
Thus, the distinctive features that set apart the benchmark DRL scheduler from the proposed scheduler are its action space, observation space, reward function, and DNN architecture.

\section{Complexity of the Considered Schedulers}\label{sectionschedulerComplexities}

{
\setlength\arrayrulewidth{1pt}
\begin{table}[!t]
\caption{Complexity of Fundamental Operations \label{basicoperationcomplexity}}
\centering
\begin{tabular}{l l | l l}
\hline
\textbf{Operations} & \textbf{Complexity} & \textbf{Operations} & \textbf{Complexity}  \\ 
\hline
$\chol(\mathbf{\Psi}'(t))$ & $M^{3}/3$ & ReLU & $1$  \\
$\mathcal{N}(\mathbf{1}_{i}^{T}\hat{\mathbf{x}}, \mathbf{1}_{i}^{T} \mathbf{\Psi} \mathbf{1}_{i})$ & $1$ & $\mathbf{\Psi}^{*}(t)^{-1}$ & $M^{3}$  \\
$\argmin_{n \in \{ 1, \cdots, N \}} \nu_{n}$ & $N$ & $\mathcal{N}(\hat{\mathbf{x}}, \mathbf{\Psi})$ & $M$  \\
Inequality & $1$ & $\VAR(\boldsymbol{u})$ & $S$  \\
Draw $\theta$ from $\mathcal{U}(0,1)$ & $1$ & $z_{c}(\mathbf{x}_s)$ & $1$ \\
\hline
\end{tabular}
\end{table}
}

We quantify the computational complexity of our considered schedulers in terms of the number of arithmetic operations they perform.
Table~\ref{basicoperationcomplexity} presents the complexity expressions for fundamental operations utilized in the algorithms.
Note that the complexity expressions for our considered schedulers pertain specifically to the complexity associated with making a scheduling decision at a single time step.

Notice that deriving an exact expression for the complexity of the Monte Carlo scheduler is not feasible because of step~\ref{step5MonteCarloalgo} of Algorithm~\ref{montecarloscheduler}.
However, we can derive expressions for both the lower and upper bound of the complexity of the Monte Carlo scheduler.
The lower bound expression pertains to the case that the inequality in step~\ref{step5MonteCarloalgo} of Algorithm~\ref{montecarloscheduler} is never satisfied.
Conversely, the upper bound expression represents the case that the aforesaid inequality is always satisfied.
The lower and upper bound complexity expressions are given by
\begin{subequations}\label{initial_montecarloComplesity}
\begin{align}
     \vartheta'_{lb} = & NS \Bigl(\frac{M^3}{3} + 8M^{3}n' + 8M^{2}n' + 4M^{2} + 2Mn'\phi \nonumber \\
    & \qquad+ 4 + M \Bigr) + N, \\
    \vartheta'_{ub} = & \ \vartheta'_{lb} + NS \Bigl(\frac{22M^3}{3} + 12M^{3}n' + 10M^{2}n' \nonumber \\
    & \qquad\qquad\quad\; + 8M^{2} + 2Mn'\phi' + M + 3 \Bigr),
\end{align}
\end{subequations}
respectively.
Here, $\phi$ and $\phi'$ represents the computational complexity of the operation ${\mathbf{f}(\chol(\hat{\mathbf{\Psi}}(t-1)) \boldsymbol{\xi}_{i} + \hat{\mathbf{x}}(t-1))}$ and ${\mathbf{g}(\chol(\mathbf{\Psi}'(t)) \boldsymbol{\xi}_{i} + \mathbf{x}'(t))}$ in Algorithm~\ref{predictionstepCQKF} and Algorithm~\ref{updatestepCQKF}, respectively.

Suppose \cite{https://doi.org/10.1049/iet-spr.2012.0085}
\begin{subequations}\label{NLDS_NLOMfunction}
\begin{align}
    \label{NLDS_M3} \mathbf{f}(\mathbf{x}(t)) & = \mathbf{x}(t) + 0.01 \left[\begin{array}{ll}
        10(x_2-x_1) \\
        28x_1 - x_2 - x_1x_3 \\
	x_1x_2 - \frac{8}{3}x_3
        \end{array} \right], \\
    \label{NLOM_equation} \mathbf{g}(\mathbf{x}(t)) & = 0.01 \mathbf{x}(t) \odot (\mathbf{1}_M - 0.5\mathbf{x}(t)),
\end{align}
\end{subequations}
where ${\mathbf{x}(t) = [x_1, x_2, \cdots , x_M]^T}$ and $\odot$ signifies the element-wise product.
Thus, $(\ref{initial_montecarloComplesity})$ becomes
\begin{subequations}\label{montecarloComplexity}
\begin{align}
    \label{LBmontecarloComplexity} \vartheta'_{lb} = & NS \Bigl(\frac{M^3}{3} + 8M^{3}n' + 12M^{2}n' + 4M^{2} \nonumber \\
        & \qquad + 18Mn' + 4 + M \Bigr) + N, \\
    \label{UBmontecarloComplexity} \vartheta'_{ub} = & \ \vartheta'_{lb} + NS \Bigl(\frac{22M^3}{3} + 12M^{3}n' + 18M^{2}n' \nonumber \\
        & \qquad\qquad\quad\; + 8M^{2} + M + 3 \Bigr),
\end{align}
\end{subequations}
respectively.
By taking into account the dominant terms in $(\ref{montecarloComplexity})$, the final complexity expression for the Monte Carlo scheduler, in terms of big-O notation, is given by
\begin{align}\label{bigOMonteCarloComplexity}
    \vartheta' & = \textcolor{black}{\mathcal{O}}(NSM^{3}n').
\end{align}

The complexity expression for the proposed scheduler is the summation of the complexities across three distinct phases:
action values generation phase, action selection phase, and training phase.
The complexity expressions for \textcolor{black}{the} first and third phase are ${\sum_{i=1}^{|\boldsymbol{l}|-1} l_{i+1} (2 l_{i} + 1)}$ and ${|\mathcal{B}| \sum_{i=1}^{|\boldsymbol{l}|-1} l_{i+1} (2 l_{i} + 1)}$, respectively, as derived in \cite{10143239}.
Here, ${\boldsymbol{l} = [l_{1}, \cdots, l_{|\boldsymbol{l}|}]^T}$ with ${l_{1} = |\boldsymbol{s}(t)|}$ and ${l_{|\boldsymbol{l}|} = |\mathcal{A}|}$, while the remaining elements of ${\boldsymbol{l}}$ are the hidden layer sizes.
Moreover, because of steps~\ref{step4proposedscheduleralgo}-~\ref{step7proposedscheduleralgo} of Algorithm~\ref{CQKF_DRLscheduler}, the complexity of the second phase falls within the range ${[3, (2 + |\mathcal{A}|)]}$.
In the case of the proposed scheduler, ${\boldsymbol{l} = [(M^{2}+M+3C), 10, (N+1)]^T}$ and ${\boldsymbol{l} = [(2M+4C), 4, (N+1)]^T}$ with respect to ${\boldsymbol{s}'}$ and ${\boldsymbol{s}^{*}}$, respectively.
Thus, the lower and upper bound complexity expressions for the proposed scheduler are 
\begin{subequations}\label{proposedschedulerComplexity}
\begin{align}
    \label{LBproposedschedulerComplexity} \vartheta^{*}_{lb} = & \ (|\mathcal{B}| + 1) \sum_{i=1}^{|\boldsymbol{l}|-1} l_{i+1} (2 l_{i} + 1) + 3 \nonumber \\
        = & \left\{\begin{array}{ll}
        (20M^{2}+20M+ 60C + 21N  & \\
        \; +31) (30N+31) + 3,& \textrm{for} \ \boldsymbol{s}',\\
        (16M+ 32C + 9N & \\
        \; +13) (30N+31) + 3,& \textrm{for} \ \boldsymbol{s}^{*},
        \end{array}\right. \\
    \label{UBproposedschedulerComplexity} \vartheta^{*}_{ub} = & \ (|\mathcal{B}| + 1) \sum_{i=1}^{|\boldsymbol{l}|-1} l_{i+1} (2 l_{i} + 1) + 2 + (N+1) \nonumber \\
        = & \ \vartheta^{*}_{lb} + N,
\end{align}
\end{subequations}
respectively. By taking into account the dominant terms in $(\ref{proposedschedulerComplexity})$, the final complexity expression for the proposed scheduler is given by
\begin{align}\label{bigOproposedschedulerComplexity}
    \!\!\!\!\! \vartheta^{*} & =\! \left\{\begin{array}{ll}
        \!\!\! \textcolor{black}{\mathcal{O}}(20M^{2}N + 21N^{2} + 20MN + 60NC),& \!\!\!\!\! \textrm{for} \ \boldsymbol{s}',\\
        \!\!\! \textcolor{black}{\mathcal{O}}(9N^{2} + 32CN + 16MN),& \!\!\!\!\! \textrm{for} \ \boldsymbol{s}^{*}.
        \end{array}\right.
\end{align}

As mentioned in Section~\ref{benchmarksche2subsection}, the working of the benchmark DRL scheduler is the same as the proposed scheduler.
Thus, the general complexity expression for the benchmark DRL scheduler is the same as the ones derived for the proposed scheduler.
However, this time ${\boldsymbol{l} = [(M+M^{2}+C), 2.5M, M, N, N]^T}$.
Thus, the lower and upper bound complexity expressions for the benchmark DRL scheduler are 
\begin{subequations}\label{benchmark2Complexity}
\begin{align}
    \label{LBbenchmark2Complexity} \vartheta^{\dag}_{lb} = & (30N+1) (5M^{3} + 10M^{2} + 5MC + 3.5M \nonumber \\
        & \qquad\qquad\;\; + 2NM + 2N^{2} + 2N) + 3, \\
    \label{UBbenchmark2Complexity} \vartheta^{\dag}_{ub} = & \  \vartheta^{\dag}_{lb} + N - 1,
\end{align}
\end{subequations}
respectively. By taking into account the dominant terms in $(\ref{benchmark2Complexity})$, the final complexity expression for the benchmark DRL scheduler is given by
\begin{align}\label{bigOScheduler2Complexity}
    \vartheta^{\dag} & = \textcolor{black}{\mathcal{O}}(5M^{3}N + 5MCN + 2MN^{2} + 2N^{3}).
\end{align}
From $(\ref{bigOMonteCarloComplexity})$, $(\ref{bigOproposedschedulerComplexity})$ and $(\ref{bigOScheduler2Complexity})$, we observe that the proposed scheduler has cubic and quadratic computational complexity with respect to ${\boldsymbol{s}'}$ and ${\boldsymbol{s}^{*}}$, while the benchmark schedulers have polynomial computational complexity.
Moreover, by taking into account $(\ref{montecarloComplexity})$, $(\ref{proposedschedulerComplexity})$, $(\ref{benchmark2Complexity})$, and $\{S,n'\} = \{100,4\}$, the complexities of the considered schedulers for various system parameter configurations are available in Table~\ref{schedulercomplexityTable}.
As can \textcolor{black}{be} seen in Table~\ref{schedulercomplexityTable}, complexities of the proposed scheduler with respect to ${\boldsymbol{s}^{*}}$ are extremely small for all the system parameter configurations.

\textcolor{black}{Lastly, the computational latency of the Cholesky decomposition operation is given by
\begin{align}\label{Chol_comp_latency}
    \vartheta_{cl} & = \frac{M^{3}}{3 \digamma},
\end{align}
where $M^{3}/3$ is the complexity of the Cholesky decomposition \cite{7853809}, while $\digamma$ denotes the floating-point operations per second (FLOPS) of an embedded processor in an edge node.
When $\digamma = \{80,200\}$ mega FLOPS, $\vartheta_{cl}$ is $\{\sim0.11,\sim0.045\}\mu$s and $\{\sim3.04,\sim1.215\}\mu$s for $M=3$ and $M=9$, respectively.
Therefore, the low computational latency of the Cholesky decomposition and the low complexity with respect to ${\boldsymbol{s}^{*}}$ in Table~\ref{schedulercomplexityTable} render CQKF-cum-DRL-based scheduler suitable for implementation on an embedded processor-based edge node.}

{
\setlength\arrayrulewidth{1pt}
\begin{table}[!t]
\caption{Complexities for Various System Parameter Configurations \label{schedulercomplexityTable}}
\centering
\begin{tabular}{@{}l l@{} l@{} l@{}}
\hline
\begin{tabular}{@{}l}
    $\{N,M,$ \\
    $\qquad C\}$
\end{tabular} &
\hspace{-2.2em} \begin{tabular}{p{1cm} l}
                    \multicolumn{2}{l}{\textbf{Proposed Scheduler}} \\
                    $\boldsymbol{s}' (\times 10^{5})$ & $\boldsymbol{s}^{*} (\times 10^{5})$
                   \end{tabular} & \hspace{-1.1em} \begin{tabular}{l}
                    \textbf{Benchmark DRL} \\
                    \textbf{Scheduler}$(\times 10^{5})$
                   \end{tabular} & \hspace{-1.1em} \begin{tabular}{l@{}}
                    \textbf{Monte Carlo} \\
                    \textbf{Scheduler}$(\times 10^{5})$
                   \end{tabular} \\
\hline
$\{3,3,2\}$ & \hspace{-2.2em} \begin{tabular}{p{1cm} l}
                    $\sim 0.5$ & $\sim 0.2$
                   \end{tabular} & $\sim 1.6$ & $\sim 5 - 11$ \\
$\{3,3,5\}$ & \hspace{-2.2em} \begin{tabular}{p{1cm} l}
                    $\sim 0.8$ & $\sim 0.3$
                   \end{tabular} & $\sim 2$ & $\sim 5 - 11$ \\
$\{5,5,2\}$ & \hspace{-2.2em} \begin{tabular}{p{1cm} l}
                    $\sim 1.5$ & $\sim 0.4$
                   \end{tabular} & $\sim 6$ & $\sim 28 - 73$ \\
\hline
\end{tabular}
\end{table}
}

\begin{figure}[!t]
\centering
\includegraphics[width=0.8\linewidth]{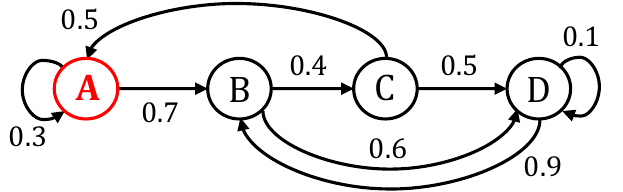}
\caption{An illustration of the SMP. Here, ${\mathcal{Q}_c = \{\textbf{A}, \textbf{B}, \textbf{C}, \textbf{D}\} , \forall c}$. Note that A client asks a query when its corresponding SMP reaches state $\textbf{A}$.}
\label{SMPfigure}
\end{figure}

\section{Simulations}\label{resultssection}

Our simulations consider \textcolor{black}{the non-linear observation function from (\ref{NLOM_equation}) and two NLSD functions: (\ref{NLDS_M3}) with $M=3$ and
\begin{subequations}\label{NLDS_M9}
\begin{align}
    \mathbf{f}(\mathbf{x}(t)) & = \mathbf{x}(t) + 0.01 \mathbf{x}_{0}, \\
    \mathbf{x}_{0} & = \left[\begin{array}{ll}
    -\frac{1}{2} b_1 x_{1} - x_{2} x_{4} + b_4 x_{4}^{2} + b_3 x_{3} x_{5} - \frac{1}{2} b_2 x_{7} \\
	-\frac{1}{2} x_{2} + x_{1} x_{4} - x_{2} x_{5} + x_{4} x_{5} - \frac{1}{4} x_{9} \\
    -\frac{1}{2} b_1 x_{3} + x_{2} x_{4} - b_4 x_{2}^{2} - b_3 x_{1} x_{5} + \frac{1}{2} b_2 x_{8} \\
    -\frac{1}{2} x_{4} - x_{2} x_{3} - x_{2} x_{5} + x_{4} x_{5} + \frac{1}{4} x_{9} \\
    -\frac{1}{2} b_5 x_{5} + \frac{1}{2} x_{2}^{2} - \frac{1}{2} x_{4}^{2} \\
    -b_6 x_{6} + x_{2} x_{9} - x_{4} x_{9} \\
    -b_1 x_{7} - 14.1 x_{1} + 2 x_{5} x_{8} - x_{4} x_{9} \\
    -b_1 x_{8} + 14.1 x_{3} - 2 x_{5} x_{7} + x_{2} x_{9} \\
    -x_{9} - 14.1 x_{2} + 14.1 x_{4} - 2 x_{2} x_{6} + 2 x_{4} x_{6} \\
    \qquad\qquad\qquad\qquad\qquad\qquad+ x_{4} x_{7} - x_{2} x_{8}
        \end{array} \right]
\end{align}
\end{subequations}
from \cite{Peter_Reiterer_1998} with $M=9$.
Notably, (\ref{NLDS_M3}) and (\ref{NLDS_M9}) lead} to NLDS with correlated states. \textcolor{black}{We assume $N=M$.}
Furthermore, MQP considered for simulations is the semi-Markov process (SMP), whose illustration is available in Fig.~\ref{SMPfigure}.
The SMP's state at time $t$ is governed by the transition matrix and the holding time distribution.
The information about the holding time distribution and queries asked by clients for the case $C=2$ is available in Table~\ref{clientcombination_table}.
Note that in Table~\ref{clientcombination_table}, $\mathfrak{C}$ refers to the SMP combinations possible at the client side, while $\eth_{i_c}$ refers to the holding time and is defined as the time duration for which the SMP stays in its state $i_c$ before transitioning.
Also, the transition matrix and the holding time distribution are unknown to the edge node.
Table~\ref{simulation_parameter_table} shows the default parameter values in our simulations, \textcolor{black}{while Table} \ref{DRLschedulerarchitecture} and \ref{LSTMarchitecture} \textcolor{black}{present} insights \textcolor{black}{into} the architecture of the online/target network and the LSTM network, respectively.

After the training phase, the performance evaluation of schedulers is performed over a duration of $2000$ time steps through ${\widehat{\MSE}_{c}(t), \forall c \in \mathscr{C}}$, and action selection frequency (ASF) metrics.

{
\setlength\arrayrulewidth{1pt}
\begin{table}[!t]
\caption{Information About Clients, for the Case $C=2$ \label{clientcombination_table}}
\centering
\begin{tabular}{@{}l l l l l@{}}
\hline
\textbf{Parameters} & \textbf{state} & $\mathfrak{C}$ & \textbf{Client-1} & \textbf{Client-2} \\
\hline
Holding & $i_{c}^\dagger$ & $1$ & Logarithmic$^{\ddagger}$ & Logarithmic  \\
time & $i_{c}^\dagger$ & $2$ & Zipf$^{\S}$ & Zipf  \\
distribution & $i_{c}^\dagger$ & $3$ & Zipf & Logarithmic  \\
 & $\textbf{A}_c$ & $\forall \mathfrak{C}$ & $\left\{\begin{array}{ll}
    \!\!\! 1,& \!\!\!\! \eth_{\textbf{A}_c} = 1, \\
    \!\!\! 0,& \!\!\!\! \textrm{otherwise}.
\end{array}\right.$ & $\left\{\begin{array}{ll}
    \!\!\! 1,& \!\!\!\! \eth_{\textbf{A}_c} = 1, \\
    \!\!\! 0,& \!\!\!\! \textrm{otherwise}.
\end{array}\right.$  \\
Query asked & $-$ & $-$ & Maximum query & Count range query \\
\hline
\multicolumn{5}{l}{$^{\dagger}$ $\{i_c \mid \forall (i_c \in \mathcal{Q}_c) \cap (i_c \neq \textbf{A}_c), \forall c\}$.}\\
\multicolumn{5}{l}{$^{\ddagger}$ $- \mho^{\eth_{i_c}} / (\eth_{i_c} \ln(1-\mho))$.}\\
\multicolumn{5}{l}{$^{\S}$ $\eth_{i_c}^{-\mho'} / \textcolor{black}{R'}(\mho')$, here $\textcolor{black}{R'}(\cdot)$ is the Riemann zeta function.}
\end{tabular}
\end{table}
}

{
\setlength\arrayrulewidth{1pt}
\begin{table}[!t]
\caption{Parameters Used in Simulations\label{simulation_parameter_table}}
\centering
\begin{tabular}{l l@{}}
\hline
\textbf{Parameters} & \textbf{Values}  \\ 
\hline
$\mathbf{\Sigma}$ & $2.5 \times 10^{-3} \mathbf{I}_M$  \\
$n'$ & $4$ \\
$S$ & $800$ \\
$[a,  b]$ & $[5, 10]$ \\
$\hat{\mathbf{\Psi}}(0)$ & $0.35\mathbf{I}_M$~\cite{https://doi.org/10.1049/iet-spr.2012.0085} \\
$\mathbf{\Sigma}'$ & $6.5 \times 10^{-4} \mathbf{I}_M$ \\
$\mho$ & $0.1$ \\
$\mho'$ & $5.35$ \\
$\{N, M\}$ \textcolor{black}{from (\ref{NLDS_M3})} & $3$~\cite{https://doi.org/10.1049/iet-spr.2012.0085} \\
$\{N, M\}$ \textcolor{black}{from (\ref{NLDS_M9})} & \textcolor{black}{$9$~\cite{Peter_Reiterer_1998}} \\
$\hat{\mathbf{x}}(0)$ \textcolor{black}{for (\ref{NLDS_M3})} & $[1.35, -3, 6]$~\cite{https://doi.org/10.1049/iet-spr.2012.0085} \\
\textcolor{black}{$\hat{\mathbf{x}}(0)$ for (\ref{NLDS_M9}) }& \textcolor{black}{$[0.01, 0, 0.01, 0, 0, 0, 0, 0, 0.01 ]$~\cite{Peter_Reiterer_1998}} \\
\textcolor{black}{$\{b_1, b_2, b_3, b_4, b_5, b_6\}$} & \textcolor{black}{$\{\frac{10}{3}, \frac{3}{5}, \frac{6}{5}, \frac{1}{5}, \frac{4}{3}, \frac{8}{3}\}$~\cite{Peter_Reiterer_1998}} \\
\hline
\end{tabular}
\end{table}
}

{
\setlength\arrayrulewidth{1pt}
\begin{table}[!t]
\caption{Online and Target Network Architecture and Parameters \label{DRLschedulerarchitecture}}
\centering
\begin{tabular}{@{}l l@{}}
\hline
\textbf{Parameters} & \textbf{Values}  \\ 
\hline
Activation function & ReLU~\cite{raghuwanshi2024neural}  \\
Optimizer & RMSProp~\cite{raghuwanshi2024neural}  \\
Learning rate & $10^{-3}$  \\
Mini-batch size $(|\mathcal{B}|)$ & $|\mathcal{A}| \times 50$    \\
Memory buffer size $(|\mathcal{E}|)$ & $|\mathcal{A}| \times 400$~\cite{nabati2021online}  \\
Threshold for global norm of \\ gradient vector $(a')$ & $5.0$~\cite{nabati2021online} \\
$\Theta_{1}, \Theta_{2}$ (initialize) & $[-0.01, 0.01]$ \\
$\epsilon$ (initial value) & $1$~\cite{raghuwanshi2024neural} \\
$\eta', \eta^{*}$ (initial value) & $0$ \\
$d, d', d^{*}, \eta'_{max}, \eta^{*}_{max}$ & $30, 0.1, 0.01, 30, 20$ \\
Transmission cost $(d^{\dag})$ & $35$ \\
Exponential discount factor $(\gamma)$ & $0.9$ \\
Output dimension & $N + 1$  \\
 & \begin{tabular}{@{}p{1.7cm} p{1.7cm}@{}}
          \hline
          $\boldsymbol{s}'$ & $\boldsymbol{s}^{*}$ \\
          \hline
          \end{tabular} \\
Number of hidden layers & \begin{tabular}{@{}p{1.7cm} p{1.7cm}@{}}
          $1$  & $1$
          \end{tabular} \\
Hidden layers dimension & \begin{tabular}{@{}p{1.7cm} p{1.7cm}@{}}
          $10$  & $4$
          \end{tabular} \\
Input dimension & \begin{tabular}{@{}p{1.7cm} p{1.7cm}@{}}
          $M^{2}+M+3C$  & $2M+4C$
          \end{tabular} \\
\hline
\end{tabular}
\end{table}
}

{
\setlength\arrayrulewidth{1pt}
\begin{table}[!t]
\caption{LSTM Network Architecture and Parameters \label{LSTMarchitecture}}
\centering
\begin{tabular}{@{}l l@{}}
\hline
\textbf{Parameters} & \textbf{Values}  \\ 
\hline
Optimizer & Adam~\cite{kingma2014adam}  \\
Learning rate & $10^{-3}$  \\
Mini-batch size $(|\mathcal{B}'|)$ & $64$    \\
Training data size $(|\mathcal{E}'|)$ & $2\times10^{5}$  \\
Output dimension & $2C$  \\
Number of hidden layers & $1$ \\
Hidden layers dimension & $64$ \\
Input dimension & $2C$ \\
Input sequence length $(l')$ & $10$ \\
$\tau^{\dag}$ & 1 \\
$t'$ (initial value) & 0 \\
\hline
\end{tabular}
\end{table}
}

{
\setlength\arrayrulewidth{1pt}
\begin{table}[!t]
\caption{Number of Sensor Transmissions \label{sensortransmissionsTable}}
\centering
\begin{tabular}{@{}l l l >{\color{black}}l >{\color{black}}l l l@{}}
\hline
$\mathfrak{C}$ & \multicolumn{4}{l}{\textbf{Proposed Scheduler}} & \textbf{Benchmark} & \textbf{Monte} \\
 & \multicolumn{2}{p{2.1cm}}{\textcolor{black}{NLSD from (\ref{NLDS_M3})}} & \multicolumn{2}{p{2cm}}{\textcolor{black}{NLSD from (\ref{NLDS_M9})}} & \textbf{DRL} & \textbf{Carlo} \\
 & $\boldsymbol{s}'$ & $\boldsymbol{s}^{*}$ & $\boldsymbol{s}'$ & $\boldsymbol{s}^{*}$ & \textbf{Scheduler} & \textbf{Scheduler} \\
\hline
$1$ & $288$ & $320$ & $545$ & $422$ & $2000$ & $410$ \\
$2$ & $304$ & $297$ & $596$ & $297$ & $2000$ & $437$ \\
$3$ & $299$ & $250$ & $426$ & $277$ & $2000$ & $427$ \\
\hline
\end{tabular}
\end{table}
}

\subsection{\textcolor{black}{Convergence Analysis of the Proposed Scheduler}}\label{subsection_convergenceAnalysis}

\begin{figure*}[!t]
\captionsetup[subfigure]{labelformat=empty}
\centering
\begin{minipage}[t]{\columnwidth}
\includegraphics[width=\linewidth]{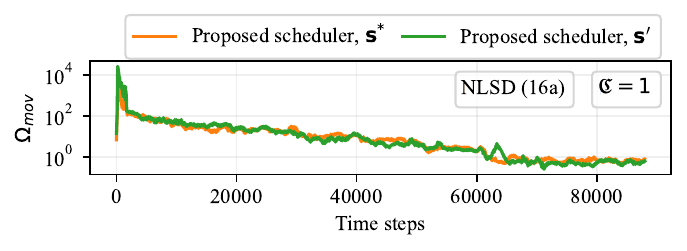}
\end{minipage}\hfill
\begin{minipage}[t]{\columnwidth}
\includegraphics[width=\linewidth]{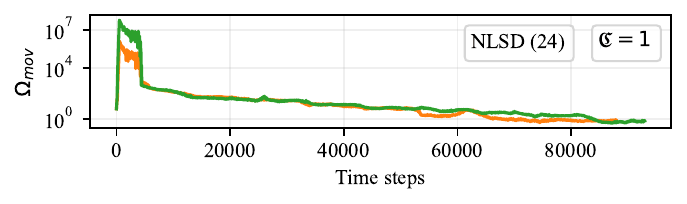}
\end{minipage}\vfill
\begin{minipage}[t]{\columnwidth}
\includegraphics[width=\linewidth]{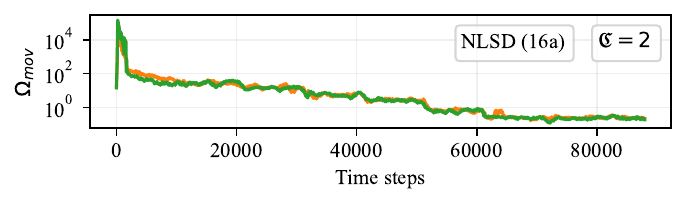}
\end{minipage}\hfill
\begin{minipage}[t]{\columnwidth}
\includegraphics[width=\linewidth]{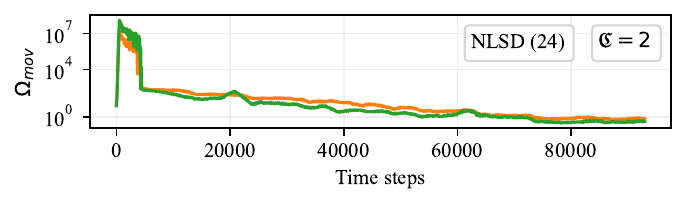}
\end{minipage}\vfill
\begin{minipage}[t]{\columnwidth}
\includegraphics[width=\linewidth]{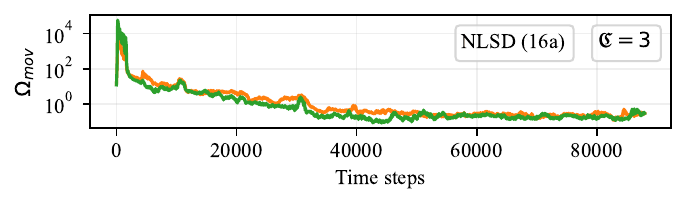}
\end{minipage}\hfill
\begin{minipage}[t]{\columnwidth}
\includegraphics[width=\linewidth]{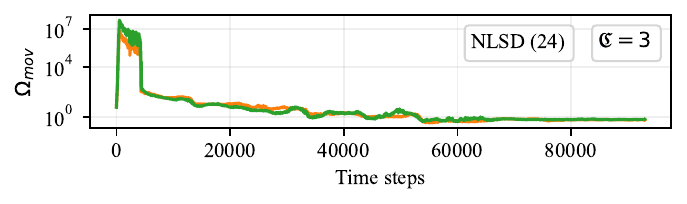}
\end{minipage}
\caption{\textcolor{black}{Time evolution of the moving average $\Omega_{mov}$, with sliding window length $51$, of $\Omega$ during the training phase of the online network for the case $C=2$ and NLSD function from (\ref{NLDS_M3}) and (\ref{NLDS_M9}).}}
\label{convergenceAnalysis_M3_M9_figure}
\end{figure*}

\textcolor{black}{Fig.~\ref{convergenceAnalysis_M3_M9_figure} illustrates that initially $\Omega_{mov}$, which is the moving average of $\Omega$, varies in the range $[\sim10^2, \sim10^7]$.
This is because the online network undergoes an exploration phase during the first $2|\mathcal{B}|$ time steps of its training phase, as mentioned in step~\ref{step37_algo6} of Algorithm~\ref{CQKF_DRLscheduler}.
After the initial exploration phase, both exploitation and exploration operations occur, and $\Omega_{mov}$ decreases with the progression of the online network's training, regardless of the NLSD function, ${\boldsymbol{s}(t)}$, and $\mathfrak{C}$.
Finally, $\Omega_{mov}$ settles at a value $<1$ for all six combinations of $\mathfrak{C}$ and NLSD functions, signifying the convergence of the online network and the proposed scheduler.}

\subsection{ASF}\label{subsection_ASF}

\begin{figure*}[!t]
\captionsetup[subfigure]{labelformat=empty}
\centering
\begin{minipage}[t]{\columnwidth}
\includegraphics[width=\linewidth]{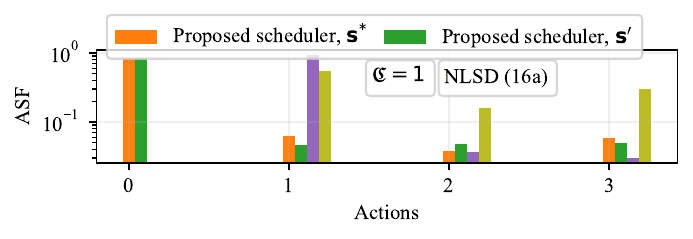}
\end{minipage}\hfill
\begin{minipage}[t]{\columnwidth}
\includegraphics[width=\linewidth]{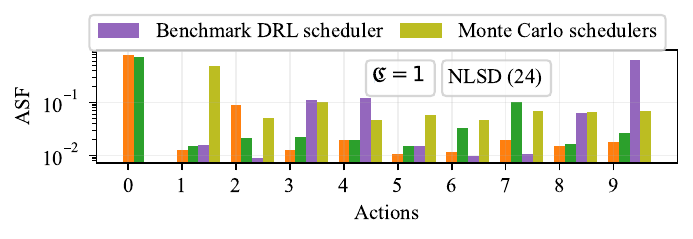}
\end{minipage}\vfill
\begin{minipage}[t]{\columnwidth}
\includegraphics[width=\linewidth]{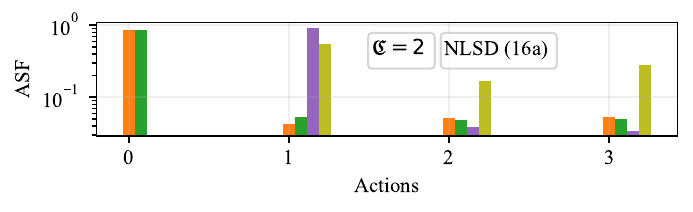}
\end{minipage}\hfill
\begin{minipage}[t]{\columnwidth}
\includegraphics[width=\linewidth]{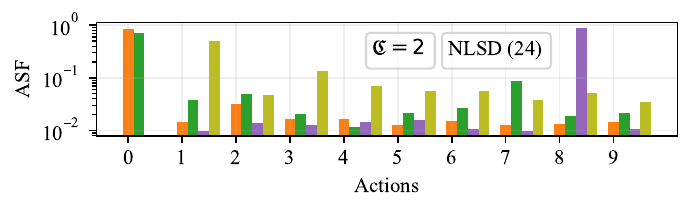}
\end{minipage}\vfill
\begin{minipage}[t]{\columnwidth}
\includegraphics[width=\linewidth]{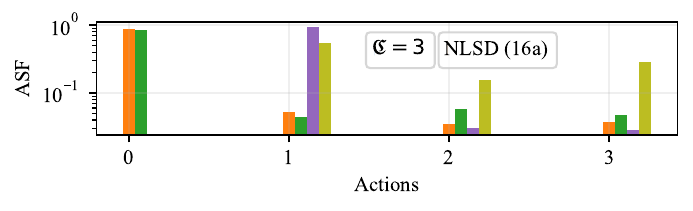}
\end{minipage}\hfill
\begin{minipage}[t]{\columnwidth}
\includegraphics[width=\linewidth]{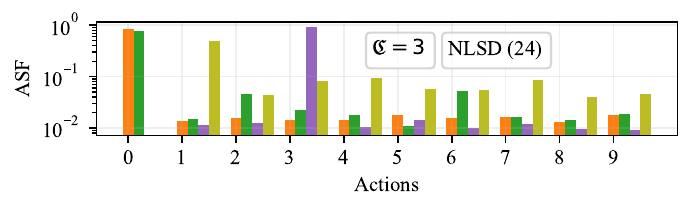}
\end{minipage}
\caption{ASFs resulting from using the considered schedulers for \textcolor{black}{the case $C=2$ and NLSD function from (\ref{NLDS_M3}) and (\ref{NLDS_M9}).}}
\label{ASF_M3_M9_figure}
\end{figure*}

\textcolor{black}{Bar plots} in Fig.~\ref{ASF_M3_M9_figure} reveal that \textcolor{black}{${\textrm{action 0}}$} is the most adopted by the proposed scheduler among all of its possible actions, \textcolor{black}{in both the cases of NLSD}.
Moreover, ASFs of most of its remaining actions are below $10^{-1}$.
This dominance of \textcolor{black}{${\textrm{action 0}}$} stems from terms ${\mathbbm{1}(p=0) r'}$ and $d^{\dag}$ incorporated in ${r(t)}$, which incentivize the scheduler to opt for \textcolor{black}{${\textrm{action 0}}$} when transmission cost becomes significantly higher than the deduction in the $\MSE$ of the query response.
Selecting \textcolor{black}{${\textrm{action 0}}$} also minimizes sensor transmissions, thereby saving sensor energy.
In contrast, benchmark schedulers predominantly select \textcolor{black}{${\textrm{action 1}}$ across all ${\mathfrak{C}}$ in the case of the NLSD function from (\ref{NLDS_M3}), while ${\textrm{action 1}}$, $3$, $8$, and $9$ in the case of the NLSD function from (\ref{NLDS_M9}).
This results in a substantial amount of energy depletion in ${\textrm{sensor 1}}$, $3$, $8$, and $9$.}
Furthermore, the proposed scheduler requires \textcolor{black}{fewer} sensor transmissions relative to the benchmark DRL scheduler, as evidenced in Table~\ref{sensortransmissionsTable}.
Note that the proposed scheduler opts for \textcolor{black}{${\textrm{action 0}}$} during \textcolor{black}{$70\% - 87\%$} of the testing phase.
Consequently, the sensor energy depletion is relatively lower compared to the benchmark DRL scheduler.

\begin{figure*}[!t]
\captionsetup[subfigure]{labelformat=empty}
\centering
\begin{minipage}[t]{\columnwidth}
\includegraphics[width=\linewidth]{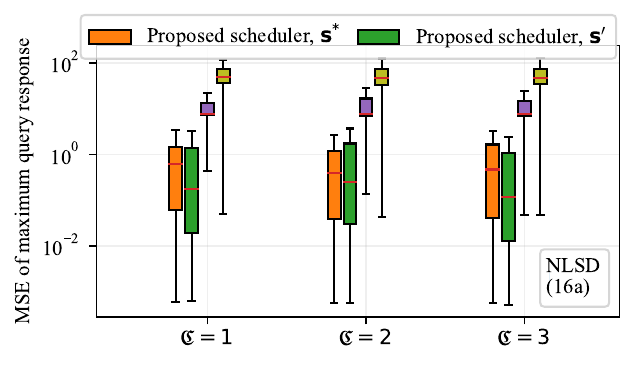}
\end{minipage}\hfill
\begin{minipage}[t]{\columnwidth}
\includegraphics[width=\linewidth]{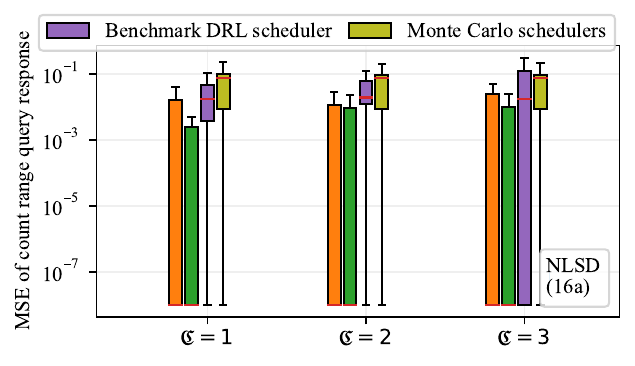}
\end{minipage}\vfill
\begin{minipage}[t]{\columnwidth}
\includegraphics[width=\linewidth]{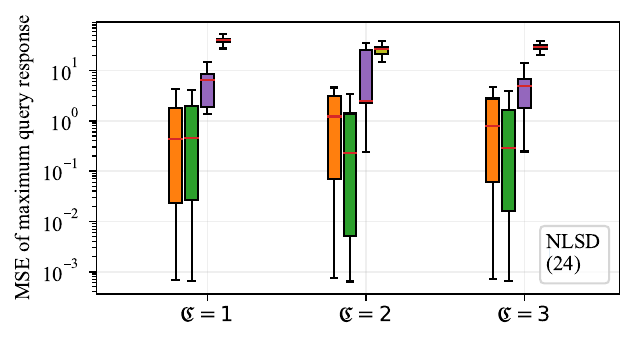}
\end{minipage}\hfill
\begin{minipage}[t]{\columnwidth}
\includegraphics[width=\linewidth]{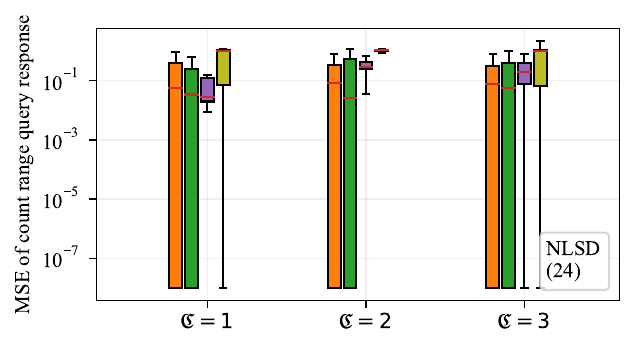}
\end{minipage}
\caption{${\widehat{\MSE}_{c}(t)}$ of the maximum \textcolor{black}{and count range query response, for the case $C=2$ and NLSD functions from (\ref{NLDS_M3}) and (\ref{NLDS_M9}),} accumulated during the run of the considered schedulers for various $\mathfrak{C}$.}
\label{max_CR_query_M3_M9_figure}
\end{figure*}

\subsection{Query Response MSE: $C=2$}\label{MSE_C2_subsection}

Since the proposed scheduler with respect to ${\boldsymbol{s}^{*}}$ and ${\boldsymbol{s}'}$ are provided with partial and full information of prior estimates, respectively, the former cannot outperform the latter in terms of ${\widehat{\MSE}_{c}(t)}$.
Fig.~\ref{max_CR_query_M3_M9_figure} \textcolor{black}{supports} this statement.
Box-plots in Fig.~\ref{max_CR_query_M3_M9_figure} \textcolor{black}{show} that there is a significant disparity in ${\widehat{\MSE}_{c}(t)}$, ${\forall c}$, obtained in the case of proposed schedulers and benchmark schedulers.
Specifically, proposed schedulers obtain a smaller ${\widehat{\MSE}_{c}(t), \forall c}$, compared to benchmark schedulers, \textcolor{black}{even as $M$ and $|\mathcal{A}|$ increase.}
Such disparity signifies the advantage of incorporating ${\MSE'_{c}(t)}$, ${\forall c}$, in \textcolor{black}{$\{\boldsymbol{s}', \boldsymbol{s}^{*}\}$ and proves the scalability of the proposed schedulers with respect to $M$ and $|\mathcal{A}|$.}

Based on the preceding discussion, it is apparent that proposed schedulers obtain a smaller ${\widehat{\MSE}_{c}(t)}$ relative to benchmark schedulers.
Furthermore, proposed schedulers accomplish this by reducing the number of sensor transmissions.
The key to \textcolor{black}{the} satisfactory performance of proposed schedulers lies in their input and reward.
Instead of feeding just the information of prior estimates, i.e., ${\{\mathbf{x}'(t), \mathbf{\Psi}'(t)\}}$, as input to the DRL scheduler, we are also feeding ${\MSE'_{c}(t), \forall c}$.
As mentioned in Section~\ref{GoS_subsection}, ${\MSE'_{c}(t)}$ informs the DRL scheduler about the out-turn of selecting \textcolor{black}{${\textrm{action 0}}$} on the $\MSE$ of the query response.
Moreover, the reward from $(\ref{rewardequation})$ prepares the DRL scheduler for timesteps with queries in a proactive manner.
This allows the DRL scheduler to select the most fruitful action that can lead to the minimization of ${\widehat{\MSE}_{c}(t)}$.
However, providing the full information of prior estimates as input, as done with the benchmark DRL scheduler and ${\boldsymbol{s}'}$, adds an extra workload of extracting the valuable information from the input to the DRL scheduler.
Because of the extra workload, the DRL scheduler might require a complex DNN architecture, which would increase the computational complexity of the scheduling operation, as shown in Table~\ref{schedulercomplexityTable}.
Contrarily, ${\boldsymbol{s}^{*}}$ makes a way for the streamlined DNN architecture.

\subsection{Query Response MSE: $C=3$}\label{MSE_C3_subsection}

\begin{figure}[!t]
\captionsetup[subfigure]{labelformat=empty}
\centering
\begin{minipage}[t]{\columnwidth}
\includegraphics[width=\linewidth]{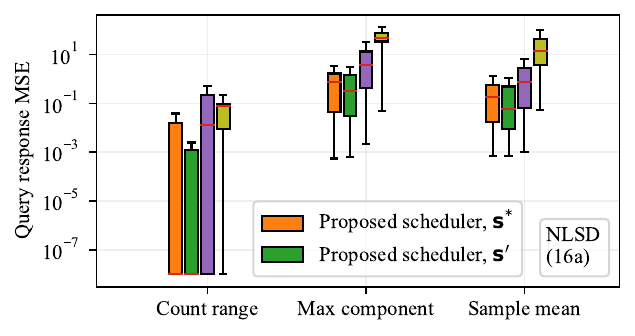}
\end{minipage}\vfill
\begin{minipage}[t]{\columnwidth}
\includegraphics[width=\linewidth]{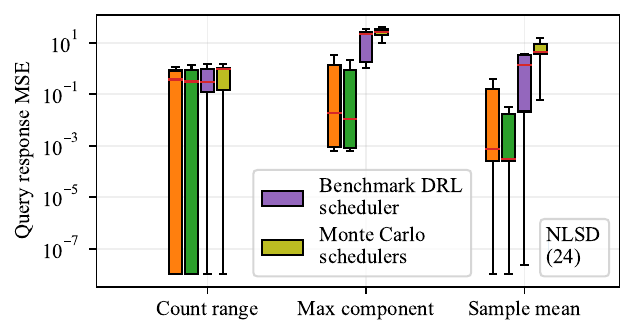}
\end{minipage}
\caption{${\widehat{\MSE}_{c}(t)}$ accumulated during the run of the considered schedulers for maximum, count range, and sample mean query responses.
\textcolor{black}{We consider the case ${C=3}$ and NLSD functions from (\ref{NLDS_M3}) and (\ref{NLDS_M9}).}
The query process on all three clients is modelled through the logarithmic distribution.
Moreover, clients 1, 2, and 3 are asking maximum, count range, and sample mean queries, respectively.}
\label{C4figure}
\end{figure}

Fig.~\ref{C4figure} considers the scenario where alongside the maximum and count range queries, an additional sample mean query is posed to the edge node by an additional client.
Fig.~\ref{C4figure} manifests that proposed schedulers obtain a smaller ${\widehat{\MSE}_{c}(t), \forall c}$, compared to benchmark schedulers.
Finally, even with an increase in \textcolor{black}{$C$, $M$, and $|\mathcal{A}|$,} the performance of the proposed schedulers is better than \textcolor{black}{the} benchmark schedulers.
\textcolor{black}{This proves the scalability of the proposed schedulers with respect to $C$, $M$, and $|\mathcal{A}|$.}

\section{Conclusion} \label{conclusionsection}

This paper introduced a GoS method tailored for IoT \textcolor{black}{sensors,} sensing NLDS.
The reporting operation is scheduled by the edge node and the phrase goal-oriented in GoS emphasizes its primary objective, which is to accurately respond to client queries regarding the NLDS state.
Through GoS, the edge node gathers partial yet insightful sensor observations to advance towards its objective.
These observations, along with a state estimator, are used to estimate the complete NLDS state, which is later employed to generate query responses.
Moreover, our findings showed that the proposed GoS yields an energy-efficient state observation from the sensor perspective.

Our work here considers only a single RL agent due to the centralized nature of the scheduling.
A promising avenue for future research would be to adapt the proposed goal-oriented sensor scheduling framework to a multi-agent RL system, such as unmanned aerial vehicle swarm where each RL agent acts as a sensor scheduler.
Meanwhile, Kalman filter-based state estimators demand information about the dynamic system model.
Thus, a second future research avenue would be to design a deep learning-based state estimate, avoiding the need for information about both the dynamic system model and the observation model.
Furthermore, a third future research avenue would be to adapt the proposed scheduling framework to a control system, where the controller performs control action based on the query response received from the edge node.

\bibliographystyle{IEEEtran}
\bibliography{IEEEabrv,references}

\begin{IEEEbiographynophoto}{Prasoon Raghuwanshi}
received the B.Tech and M.Tech degrees in Electronics $\&$ Communication Engineering from the National Institute of Technology Hamirpur, India, in 2020, and is currently pursuing the D.Sc. degree in Communications Engineering from the University of Oulu, Finland. He was a visiting researcher at the Indian Institute of Technology Indore, India, in $2024$. He is a grantee of the Nokia Scholarship, Riitta ja Jorma J. Takasen säätiön grant, and Oulun yliopiston tukisäätiö grant. His research interests include random access protocols for IoT networks, goal-oriented communications, deep reinforcement learning, and TinyML.
\end{IEEEbiographynophoto}

\vspace{11pt}

\begin{IEEEbiographynophoto}{Onel Luis Alcaraz López}
(Senior Member, IEEE) received the B.Sc. degree (Hons.) from the Central University of Las Villas, Cuba, in 2013, the M.Sc. degree from the Federal University of Paraná, Brazil, in 2017, and the D.Sc. degree (Hons.) in electrical engineering from the University of Oulu, Finland. In 2013 and 2015, he worked as a Telematics Specialist with Cuban Telecommunications Company (ETECSA). In 2020, he was a Post-Doctoral Researcher in a joint project between the University of Oulu and Nokia Oulu, Finland. He was on a six-month research visit to Rice University and the University of Houston, TX, USA, in 2024. He is currently an Associate Professor (tenure track) in sustainable wireless communications engineering with the Centre for Wireless Communications (CWC), Oulu, Finland. His research interests include sustainable IoT, energy harvesting, wireless RF energy transfer, wireless connectivity, machine-type communications, and cellular-enabled sensing and positioning systems. He is also a Collaborator to the 2016 Research Award given by Cuban Academy of Sciences, a co-recipient of the 2019 and 2023 IEEE European Conference on Networks and Communications (EuCNC) Best Student Paper Award, and a recipient of both the 2020 Best Doctoral Thesis Award granted by Academic Engineers and Architects in Finland TEK and Tekniska Föreningen i Finland TFiF in 2021 and the 2022 Young Researcher Award in the field of technology in Finland. He is currently an Associate Editor of IEEE Transactions on Communications, IEEE Wireless Communications Letters, and IEEE Communications Letters.
\end{IEEEbiographynophoto}

\vspace{11pt}

\begin{IEEEbiographynophoto}{I-Hong Hou}
(Senior Member, IEEE) is a Professor in the ECE Department of the Texas A$\&$M University. He received his Ph.D. from the Computer Science Department of the University of Illinois at Urbana-Champaign. His research interests include wireless networks, edge/cloud computing, and reinforcement learning. His work has received the 2025 IEEE Communications Society William R. Bennett Prize, the Best Paper Award from ACM MobiHoc 2017 and ACM MobiHoc 2020, and Best Student Paper Award from WiOpt 2017. He has also received the C.W. Gear Outstanding Graduate Student Award from the University of Illinois at Urbana-Champaign, and the Silver Prize in the Asian Pacific Mathematics Olympiad.
\end{IEEEbiographynophoto}

\vspace{11pt}

\begin{IEEEbiographynophoto}{Vimal Bhatia}
(SM’12, FIETE, FIEI, FOSI) is currently working as a Professor (HAG) with the Indian Institute of Technology (IIT) Indore, India, Docent at University of Oulu, and is an adjunct faculty at IIT Delhi and IIIT Delhi, India. He received Ph.D. degree from Institute for Digital Communications with The University of Edinburgh, Edinburgh, U.K., in 2006. During Ph.D. he also received the IEE fellowship for collaborative research at the  Department of Systems and Computer Engineering, Carleton University, Canada, and is Young Faculty Research Fellow from MeitY, Govt of India. He is also a recipient of the Nokia Visiting Professor Finland (2024), Prof. S. N. Mitra Memorial Award (2024), Prof. SVC Aiya Memorial Award (2019), National Slovak Fellowship, and European Mobility Scheme for Senior Researcher at Czech Republic. He has worked with various IT companies for over 11 years both in India and the UK. He is a PI/co-PI/coordinator for external projects with funding of over USD 35 million from MeitY, DST, UKIERI, MoE, AKA, IUSSTF and KPMG. He has more than 480 peer reviewed publications, and has filed 17 patents (with 14 granted). He has supervised 29 awarded PhD thesis. His research interests are in the broader areas of communications, non-Gaussian non-parametric signal processing, machine/deep learning with applications to communications and photonics. He is a reviewer for IEEE, Elsevier, Wiley, Springer, and IET. He is currently Senior Member of IEEE, Fellow IETE and certified SCRUM Master. He was also the General Co-Chair for IEEE ANTS 2018, and General Vice-Chair for IEEE ANTS 2017. He has delivered many talks, tutorials and conducted faculty development programs for the World Bank’s NPIU TEQIP-III, invited talk at WWRF46-Paris and others. He is currently Associate Editor for IETE Technical Review, Frontiers in Communications and Networks, Frontiers in Signal Processing, IEEE Transactions on Green Communications and Networking, and IEEE Wireless Communications Letters. He is currently member Steering Committee for IEEE ANTS. He has been mentioned among the World's Top 2$\%$ Scientists by the Stanford University since 2021.
\end{IEEEbiographynophoto}

\vspace{11pt}

\begin{IEEEbiographynophoto}{Matti Latva-aho}
received the M.Sc., Lic.Tech. and Dr. Tech (Hons.) degrees in Electrical Engineering from the University of Oulu, Finland in 1992, 1996 and 1998, respectively. From 1992 to 1993, he was a Research Engineer at Nokia Mobile Phones, Oulu, Finland after which he joined the Centre for Wireless Communications (CWC) at the University of Oulu. Prof. Latva-aho was Director of CWC during the years 1998-2006 and Head of the Department for Communication Engineering until August 2014. Currently, he is a Professor at the University of Oulu on wireless communications and Vice Rector for Research. He is also a Global Research Fellow with Tokyo University. His research interests are related to mobile broadband communication systems and currently, his group focuses on 6G systems research. Prof. Latva-aho has published over 500 conference or journal papers in the field of wireless communications. He received Nokia Foundation Award in 2015 for his achievements in mobile communications research.
\end{IEEEbiographynophoto}

\end{document}